\documentclass[12pt,a4paper]{article}
\usepackage[utf8]{inputenc}
\usepackage{amsmath}
\usepackage{amsfonts} 
\usepackage{mathtools} 
\usepackage{amssymb} 
\usepackage{subfigure} 
\usepackage{graphicx}
\usepackage{cite} 
\usepackage{mathrsfs}
\usepackage{hyperref}

\graphicspath{{Figs/}}

\newcommand{\abs}[1]{|#1|} 
\newcommand{\ABS}[1]{\left|#1\right|} 
\newcommand{\re}[1]{\text{Re}\left(#1\right)}
\newcommand{\im}[1]{\text{Im}\left(#1\right)}
\newcommand{\refEQ}[1]{eq.\,\eqref{#1}} 
\newcommand{\refEQS}[1]{eqs.\,\eqref{#1}} 

\newcommand{\BRA}[1]{\langle #1 |}
\newcommand{\KET}[1]{| #1 \rangle}

\newcommand{\BRAoKET}[3]{\BRA{#1}#2\KET{#3}}
\newcommand{\NeutralMeson}{M}
\newcommand{\NM}{\NeutralMeson^0}\newcommand{\NMb}{\bar\NeutralMeson^0}
\newcommand{\NMH}{\NeutralMeson_{H}}\newcommand{\NML}{\NeutralMeson_{L}}
\newcommand{\KETNM}{\KET{\NM}}
\newcommand{\KETNMb}{\KET{\NMb}}
\newcommand{\KETNMH}{\KET{\NMH}}
\newcommand{\KETNML}{\KET{\NML}}
\newcommand{\NMHeff}{{\mathbf H}}\newcommand{\NMM}{{\mathbf M}}\newcommand{\NMG}{{\mathbf \Gamma}}
\newcommand{\DMU}{\Delta\mu }\newcommand{\DM}{\Delta M}\newcommand{\DG}{\Delta\Gamma}
\newcommand{\G}{\Gamma}
\newcommand{\GBs}{\G_{B_s}}
\newcommand{\DMBs}{\DM_{B_s}}
\newcommand{\DGBs}{\DG_{B_s}}
\newcommand{\MU}{\mu}\newcommand{\MUH}{\MU_{H}}\newcommand{\MUL}{\MU_{L}}

\newcommand{\pH}{p_{H}}\newcommand{\pL}{p_{L}}\newcommand{\qH}{q_{H}}\newcommand{\qL}{q_{L}}
\newcommand{\AmpBase}{A}
\newcommand{\Af}{\AmpBase_{f}}\newcommand{\Abf}{\bar\AmpBase_{f}}
\newcommand{\CcDM}{\cos\left(\DM\, t\right)}\newcommand{\ChDG}{\cosh\left(\frac{\DG}{2}t\right)}
\newcommand{\ScDM}{\sin\left(\DM\, t\right)}\newcommand{\ShDG}{\sinh\left(\frac{\DG}{2}t\right)}
\newcommand{\CcDMBs}{\cos\left(\DMBs\, t\right)}\newcommand{\ChDGBs}{\cosh\left(\frac{\DGBs}{2}t\right)}
\newcommand{\ScDMBs}{\sin\left(\DMBs\, t\right)}\newcommand{\ShDGBs}{\sinh\left(\frac{\DGBs}{2}t\right)}

\newcommand{\Lf}{\lambda_{f}}
\newcommand{\Cf}{{\rm C}_{f}}
\newcommand{\Sf}{{\rm S}_{f}}
\newcommand{\Rf}{{\rm R}_{f}}

\newcommand{\Lkk}{\lambda_{B_s\to KK}}
\newcommand{\Ckk}{{\rm C}_{KK}}
\newcommand{\Skk}{{\rm S}_{KK}}
\newcommand{\Rkk}{{\rm R}_{KK}}

\newcommand{\ACP}[1]{A_{CP}^{#1}}
\newcommand{\CLHCb}{C_{K^+K^-}}
\newcommand{\SLHCb}{S_{K^+K^-}}
\newcommand{\RLHCb}{A_{K^+K^-}^{\Delta\Gamma}}
\newcommand{\thBs}{\theta_{B_s}}

\newcommand{\dBs}{\delta_{B_s}}

\renewcommand{\GBs}{\G}\renewcommand{\DGBs}{\DG}\renewcommand{\DMBs}{\DM}
\renewcommand{\dBs}{\delta}\renewcommand{\thBs}{\theta}
\newcommand{\COfont}{\mathscr }
\newcommand{\COC}[3]{\COfont C_{#1}[#2,#3]}
\newcommand{\COS}[3]{\COfont S_{#1}[#2,#3]}
\newcommand{\COCcBASE}{\COfont C_{c}}
\newcommand{\COChBASE}{\COfont C_{h}}
\newcommand{\COScBASE}{\COfont S_{c}}
\newcommand{\COShBASE}{\COfont S_{h}}
\newcommand{\COCc}[2]{\COC{c}{#1}{#2}}
\newcommand{\COCh}[2]{\COC{h}{#1}{#2}}
\newcommand{\COSc}[2]{\COS{c}{#1}{#2}}
\newcommand{\COSh}[2]{\COS{h}{#1}{#2}}
\newcommand{\SUMCh}{\COChBASE^\Sigma}
\newcommand{\SUMCc}{\COCcBASE^\Sigma}
\newcommand{\SUMSh}{\COShBASE^\Sigma}
\newcommand{\SUMSc}{\COScBASE^\Sigma}
\newcommand{\DIFCh}{\COChBASE^\Delta}
\newcommand{\DIFCc}{\COCcBASE^\Delta}
\newcommand{\DIFSh}{\COShBASE^\Delta}
\newcommand{\DIFSc}{\COScBASE^\Delta}

\newcommand{\Bs}{B^0_s}
\newcommand{\Bsb}{\bar B^0_s}
\newcommand{\BBsmix}{$\Bs$--$\Bsb$}

\newcounter{notas}
\graphicspath{{Figs/}}

\begin{document}

\hfill\begin{minipage}[r]{0.3\textwidth}\begin{flushright}  CFTP/19-004\\    IFIC/19-15 \end{flushright} \end{minipage}

\begin{center}

\vspace{0.50cm}

{\large \bf {CPT Violation in $\boldsymbol{B^0_s}$--$\boldsymbol{\bar B^0_s}$ mixing and the measurement of CP Violation in $\boldsymbol{B_s\to K^+K^-}$}}

\vspace{0.50cm}

Francisco J. Botella $^{a,}$\footnote{\texttt{Francisco.J.Botella@uv.es}},
Miguel Nebot $^{b,}$\footnote{\texttt{miguel.r.nebot.gomez@tecnico.ulisboa.pt}}

\end{center}
\vspace{0.50cm}
\begin{flushleft}
\emph{$^b$ Departament de F\`\i sica Te\`orica and Instituto de F\' \i sica Corpuscular (IFIC),\\
\quad Universitat de Val\`encia -- CSIC, E-46100 Valencia, Spain.}\\
\emph{$^a$ CFTP, Instituto Superior T\' ecnico, U. de Lisboa,\\ 
\quad Av. Rovisco Pais 1, P-1049-001 Lisboa, Portugal.} 
\end{flushleft}
%

\begin{abstract}
\noindent A simple analysis of time-dependent $B_s\to K^+K^-$ transitions, based on recent results from the LHCb experiment, is presented. The benefits of adopting a fully consistent theoretical description of the $B^0_s$--$\bar B^0_s$ mixing are stressed. It is shown that bounds on CPT violation in the $B^0_s$--$\bar B^0_s$ system can be consistently obtained and that direct CP violation in $B_s\to K^+K^-$ can be robustly established, even in the presence of CPT violation in the mixing.
\end{abstract}


\section{Introduction\label{SEC:intro}}
The neutral meson systems $K^0$--$\bar K^0$, $D^0$--$\bar D^0$, $B_d^0$--$\bar B_d^0$ and $B_s^0$--$\bar B_s^0$, are privileged systems for the study of CP, T and CPT violation. After the significant progress achieved more than a decade ago at the Tevatron \cite{Lewis:2014cka}, the exploration of the $B_s^0$--$\bar B_s^0$ system has flourished with the spectacular performance of the LHCb experiment \cite{Borissov:2013yha}.\\ 
In a recent work \cite{Aaij:2018tfw}, the LHCb collaboration has analysed time-dependent $B_s\to K^+ K^-$ processes, including in particular the CP violating asymmetry
\begin{equation}\label{eq:CPasym:00}
\ACP{B_s\to KK}(t)\equiv\frac{\Gamma(\Bsb(t)\to K^+K^-)-\Gamma(\Bs(t)\to K^+K^-)}{\Gamma(\Bsb(t)\to K^+K^-)+\Gamma(\Bs(t)\to K^+K^-)}\ .
\end{equation}
$B_s\to KK$ transitions are an interesting window into CP violation and New Physics, e.g. \cite{Fleischer:1999pa,Fleischer:2010ib,Ciuchini:2012gd}.
A central result of the analysis in \cite{Aaij:2018tfw} is a fit of the available data to the following functional form:
\begin{equation}\label{eq:CPasym:LHCb:00}
\ACP{B_s\to KK}(t)=\frac{-\CLHCb\CcDM+\SLHCb\ScDM}{\ChDG+\RLHCb\ShDG}\ ,
\end{equation}
with $\CLHCb$, $\SLHCb$ and $\RLHCb$ independent parameters. The following values are obtained:
\begin{equation}\label{eq::LHCb:pars:00}
\CLHCb=0.20\pm 0.06\pm 0.02,\ \SLHCb=0.18\pm 0.06\pm 0.02,\ \RLHCb=-0.79\pm 0.07\pm 0.10\,.
\end{equation}
While LHCb has measured previously $\CLHCb$ and $\SLHCb$ \cite{Aaij:2013tna}, the value of $\RLHCb$ represents the first experimental result for this observable. This third measurement offered LHCb the possibility of performing a consistency check. LHCb obtained 
\begin{equation}\label{eq:sum:rule:exp}
\abs{\CLHCb}^{2}+\abs{\SLHCb}^{2}+\abs{\RLHCb}^{2}=0.70\pm 0.20\,, 
\end{equation}
a result consistent with the well-known constraint
\begin{equation}\label{eq:CPT:sum:rule}
\abs{\CLHCb}^{2}+\abs{\SLHCb}^{2}+\abs{\RLHCb}^{2}=1\,.  
\end{equation}
Unfortunately, this consistency check becomes problematic, as we will see, as soon as one interprets \refEQ{eq:sum:rule:exp} as a potential $1.5\sigma$ discrepancy from \refEQ{eq:CPT:sum:rule}, as has been done, for example, in \cite{He:2017fln}, even calling \refEQ{eq:CPT:sum:rule} ``the CPT sum rule''. 
Also from the experimental fit, the LHCb collaboration has concluded that there is, in this channel, CP violation at $4\sigma$: there is a $4\sigma$ significance in the deviation of $(\CLHCb,\SLHCb,\RLHCb)$ from the value $(0,0,\pm 1)$. 
In this paper we recall that performing a fit assuming independent parameters $\CLHCb$, $\SLHCb$ and $\RLHCb$ together with \refEQ{eq:CPasym:LHCb:00} is a correct procedure if one is interested in getting the measurement of these quantities and a consistency check. But one must stress that this procedure is not fully consistent from the theoretical point of view and therefore it is not suited to perform a theoretical analysis including CPT violation. 
We show that a more general analysis might be able to provide more robust results on two respects: 
\begin{itemize}
\item[(i)]  CP violation can be established even if CPT violation in \BBsmix\ mixing is considered,
\item[(ii)] bounds on CPT violation in \BBsmix\ mixing can be obtained.
\end{itemize}
The discussion is organised as follows. In section \ref{SEC:gen}, the Weisskopf-Wigner approximation for the description of the time evolution of neutral meson systems is briefly revisited, including the general expressions for the time-dependent rates and paying special attention to the possibility of CPT violation. Although a proper and complete analysis can only be performed with the actual data, in section \ref{SEC:data} we discuss the simple procedure that we adopt, based on the available LHCb analysis, in order to have some usable ``experimental-like'' input. A consistent general analysis of this ``experimental-like'' input is presented in Section \ref{SEC:analysis}.

\section{Generalities\label{SEC:gen}}
In the Weisskopf-Wigner approximation \cite{Branco:1999fs}, the time evolution of states in the neutral meson-antimeson space $\{\KET{M^0},\KET{\bar M^0}\}$ is controlled by the $2\times 2$ effective hamiltonian $\NMHeff$:
\begin{equation}
\NMHeff=\NMM-\frac{i}{2}\NMG,\quad \NMM=\NMM^\dagger,\ \NMG=\NMG^\dagger,\qquad 
i\hbar\frac{d}{dt}\KET{\Psi(t)}=\NMHeff\KET{\Psi(t)}\,.
\end{equation}
The eigenstates of $\NMHeff$ (labelled ``$H$'' and ``$L$'' for \emph{heavy} and \emph{light}, most appropriate for the $B_d$, $B_s$ systems) read:
\begin{alignat}{3}
&\NMHeff\KETNMH =\ && \MUH\KETNMH,\qquad && \KETNMH = \pH\KETNM+\qH\KETNMb,\\
&\NMHeff\KETNML =\ && \MUL\KETNML,\qquad && \KETNML = \pL\KETNM-\qL\KETNMb,
\end{alignat}
with the eigenvalues given by\footnote{In the $B_d$ and $B_s$ systems, by construction, $\DM>0$ while the sign of $\DG$ could be positive or negative.}
\begin{equation}
\MU =\frac{\MUH+\MUL}{2}\equiv M-\frac{i}{2}\Gamma,\quad \DMU =\MUH-\MUL \equiv \DM-\frac{i}{2}\DG.
\end{equation}
It is customary to introduce parameters $\theta$ and $q/p$:
\begin{equation}
\theta \equiv\frac{\NMHeff_{22}-\NMHeff_{11}}{\DMU}\,,\quad \left(\frac{q}{p}\right)^{2}\equiv\frac{\NMHeff_{21}}{\NMHeff_{12}}\,,
\end{equation}
such that
\begin{equation}
\frac{\qH}{\pH}=\frac{q}{p}\sqrt{\frac{1+\theta}{1-\theta}}\ ,\qquad
\frac{\qL}{\pL}=\frac{q}{p}\sqrt{\frac{1-\theta}{1+\theta}}\ .
\end{equation}
It is also customary to define\footnote{The discussion of this section applies to the different neutral meson systems; $\DM$, $\DG$, $q/p$, $\delta$ and $\theta$ are, of course, specific of each case: it is understood in the following sections that we refer to those quantities for the $B^0_s$--$\bar B^0_s$ system.}
\begin{equation}
\delta\equiv\frac{1-|q/p|^2}{1+|q/p|^2}\ .
\end{equation}
$\theta\neq 0$ gives CP and CPT violation, while $\delta\neq 0$ gives CP and T violation.\\  
The hamiltonian, incorporating CPT violation in the mixing, can be compactly written \cite{Silva:2000db,AlvarezGaume:1998yr} as
\begin{equation}
\NMHeff=
\begin{pmatrix}
\MU -\frac{\DMU}{2}{\theta} & \frac{p}{q}\frac{\DMU}{2}\sqrt{1-{\theta}^{2}} \\ 
\frac{q}{p}\frac{\DMU}{2}\sqrt{1-{\theta}^{2}} & \MU +\frac{\DMU}{2}{\theta}
\end{pmatrix}\,.
\end{equation}
In order to include the decays of the neutral mesons in the description, for a final state $\KET{f}$ we have the decay amplitudes
\begin{equation}
\Af=\BRAoKET{f}{\mathcal T}{\NM},\quad \Abf=\BRAoKET{f}{\mathcal T}{\NMb}\,.
\end{equation}
Note that not imposing any assumption on the complex parameters $\Af$ and $\Abf$ implies also that we are including any potential CPT violation in the decay. 
One can then define
\begin{equation}\label{eq:Lf:00}
\Lf\equiv\frac{q}{p}\frac{\Abf}{\Af},
\end{equation}
and
\begin{equation}\label{eq:CfSfRf:00}
\Cf\equiv\frac{1-\abs{\Lf}^2}{1+\abs{\Lf}^2},\quad \Rf\equiv\frac{2\re{\Lf}}{1+\abs{\Lf}^2},\quad\Sf\equiv\frac{2\im{\Lf}}{1+\abs{\Lf}^2}\,.
\end{equation}
Notice that:
\begin{itemize}
\item[(1)] the definition of $\Lf$ in \refEQ{eq:Lf:00} is completely independent of $\theta$, i.e. $\Lf$ is defined without regard to the presence of CPT violation in the mixing. Therefore $\Lf$ is not related to $\qL$, $\pL$ or $\qH$, $\pH$, it is related to $q/p$ as defined in \refEQ{eq:Lf:00}.
\item[(2)] By construction $\Cf$, $\Sf$, $\Rf$ verify
\begin{equation}
\Cf^2+\Sf^2+\Rf^2=1\,.
\end{equation}
\end{itemize}
It is important to stress that there are only \emph{two} independent parameters in the complex quantity $\Lf$, and that there are no additional implicit assumptions concerning the decays (for example, assuming the dominance or not of some weak amplitude, etc); on that respect, $\Lf$ already provides the most general parametrisation. In addition, for a CP eigenstate $f$, $\Lf\neq \pm 1$ -- that is $\Cf\neq 0$ and $\Sf\neq 0$ --, signals CP violation.\\ 
The time-dependent rates for the evolution of initial meson and antimeson states which decay after time $t$  into $f$, including CPT violation, are:
\begin{multline}\label{eq:evol:NM:00}
\Gamma(\NM(t)\to f)=\ABS{\BRAoKET{f}{\mathcal T}{\NM(t)}}^2=\\ e^{-\Gamma t}\,\Gamma_f
\left\{\begin{matrix}
\COCh{\NM}{f}\ChDG+\COCc{\NM}{f}\CcDM \\ +\COSh{\NM}{f}\ShDG +\COSc{\NM}{f}\ScDM
\end{matrix}\right\},
\end{multline}
and
\begin{multline}\label{eq:evol:NMb:00}
\Gamma(\NMb(t)\to f)=\ABS{\BRAoKET{f}{\mathcal T}{\NMb(t)}}^2=\\ e^{-\Gamma t}\,\Gamma_f
\left\{\begin{matrix}
\COCh{\NMb}{f}\ChDG+\COCc{\NMb}{f}\CcDM \\ +\COSh{\NMb}{f}\ShDG +\COSc{\NMb}{f}\ScDM
\end{matrix}\right\},
\end{multline}
with $\Gamma_f\equiv(\abs{\Af}^2+\abs{\Abf}^2)/2$. The coefficients in \refEQS{eq:evol:NM:00}-\eqref{eq:evol:NMb:00} depend only on $\Cf$, $\Sf$, $\Rf$, $\theta$ and $\delta$; detailed expressions are given in appendix \ref{APP:gen}. It is clear that measurements of the time dependent rates in \refEQS{eq:evol:NM:00}--\eqref{eq:evol:NMb:00} can be sensitive to the presence of $\theta\neq 0$ in these coefficients, as discussed for example in \cite{Bernabeu:2016sgz}, where improved bounds on $\re{\theta}$ in the $B_d$ system were obtained (see also \cite{BABAR:2016tjc,Karan:2017coa}). This can be extended to scenarios with additional sources of CPT violation as done for example in \cite{Bernabeu:2016kva,vanTilburg:2014dka,Aaij:2016mos,Roberts:2017tmo}.\\ 
For a CP eigenstate $f$, the general form of the CP asymmetry $\ACP{M\to f}(t)$ in \refEQ{eq:CPasym:00} is
\begin{equation}\label{eq:CPasym:gen:00}
\ACP{M\to f}(t)=\frac{\DIFCh\ChDG+\DIFSh\ShDG+\DIFCc\CcDM+\DIFSc\ScDM}{\SUMCh\ChDG+\SUMSh\ShDG+\SUMCc\CcDM+\SUMSc\ScDM}
\end{equation}
with 
\begin{equation}\label{eq:coefs:sum:dif:00}
\DIFCh\equiv\COCh{+}{f}-\COCh{-}{f},\quad \SUMCh\equiv\COCh{+}{f}+\COCh{-}{f},
\end{equation}
where $\COCh{+}{f}=\COCh{\NMb}{f}$, $\COCh{-}{f}=\COCh{\NM}{f}$, and similarly for $\COChBASE\to\COShBASE,\COCcBASE,\COScBASE$; in the present case, $\NM=\Bs$, $\NMb=\Bsb$ and $f=K^+K^-$.  For $\delta=0$, which is an excellent approximation for both $B_q$--$\bar B_q$ systems, the expansion to linear order in $\theta$ (with no loss of generality in the subsequent discussion), gives
\begin{alignat}{3}
\nonumber
&\DIFCh=2\Rf\re{\theta},\qquad &&\DIFCc=-2\Cf-2\Rf\re{\theta},\\
\label{eq:DIF:lin:02}
&\DIFSh=-2\re{\theta},\qquad &&\DIFSc=2(\im{\theta}-\Sf),
\end{alignat}
\begin{alignat}{3}
\nonumber
&\SUMCh=2(1-\Sf\im{\theta}),\qquad &&\SUMCc=2\Sf\im{\theta},\\
\label{eq:SUM:lin:02}
&\SUMSh=2(\Cf\re{\theta}-\Rf),\qquad &&\SUMSc=-2\Cf\im{\theta}.
\end{alignat}
Notice that in general, with $\theta\neq 0$, all four time dependences $\ChDG$, $\ShDG$, $\CcDM$ and $\ScDM$, enter both the numerator and the denominator of $\ACP{M\to f}(t)$. It is also natural that, since $\theta\neq 0$ gives not only CPT violation but also CP violation, even for $\Lf=\pm 1$, $\ACP{M\to f}(t)\neq 0$ for $\theta\neq 0$.\\ 
Furthermore, for $\theta=0$, $\ACP{M\to f}(t)$ reduces to\footnote{Notice the sign differences $\Sf\to-\SLHCb$, $\Rf\to-\RLHCb$ among \refEQ{eq:CPasym:noCPT:00} and \refEQ{eq:CPasym:LHCb:00}: although their definition in terms of $\Lkk$ is irrelevant for the analysis in \cite{Aaij:2018tfw}, it includes an additional $-$ sign. Equation \eqref{eq:CPasym:noCPT:00} agrees with the corresponding expressions in \cite{Branco:1999fs}, e.g. (9.14) and (9.15).}
\begin{equation}\label{eq:CPasym:noCPT:00}
\ACP{M\to f}(t)=\frac{-\Cf\CcDM-\Sf\ScDM}{\ChDG-\Rf\ShDG},
\end{equation}
with, as is to be stressed once again, $\Cf^2+\Sf^2+\Rf^2=1$.\\
It is then clear that the form of $\ACP{M\to f}(t)$ in \refEQ{eq:CPasym:LHCb:00} used for example in \cite{He:2017fln} to analyze CPT violations:
\begin{itemize}
\item does not correspond to the general case with $\theta\neq 0$,
\item neither does it correspond to the CPT invariant case with $\theta=0$, since all three $\CLHCb$, $\SLHCb$ and $\RLHCb$ are fitted as free parameters.
\end{itemize}
In the following sections we illustrate the expected outcome of a general and consistent analysis. Of course, a complete analysis should use the actual experimental data: in the next section we discuss the generation and validation of reasonable ``LHCb like'' data, used then in the analysis of section \ref{SEC:analysis}.

\section{``LHCb like'' data\label{SEC:data}}
\begin{figure}[htb!]
\begin{center}
\subfigure[$\SLHCb$ vs. $\RLHCb$.\label{sFIG:S:R}]{\includegraphics[height=0.28\textwidth]{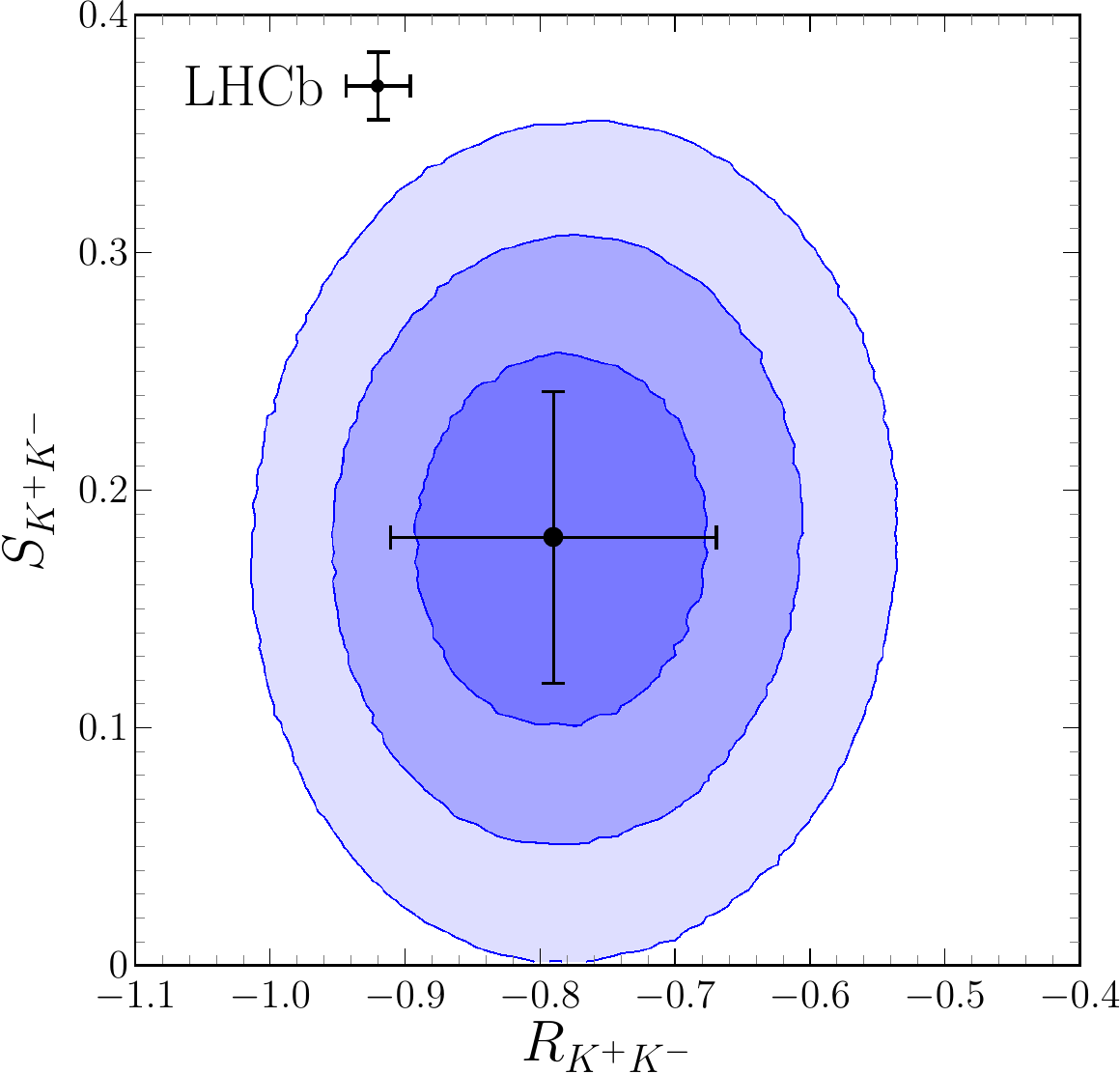}}\quad
\subfigure[$\CLHCb$ vs. $\RLHCb$.\label{sFIG:C:R}]{\includegraphics[height=0.28\textwidth]{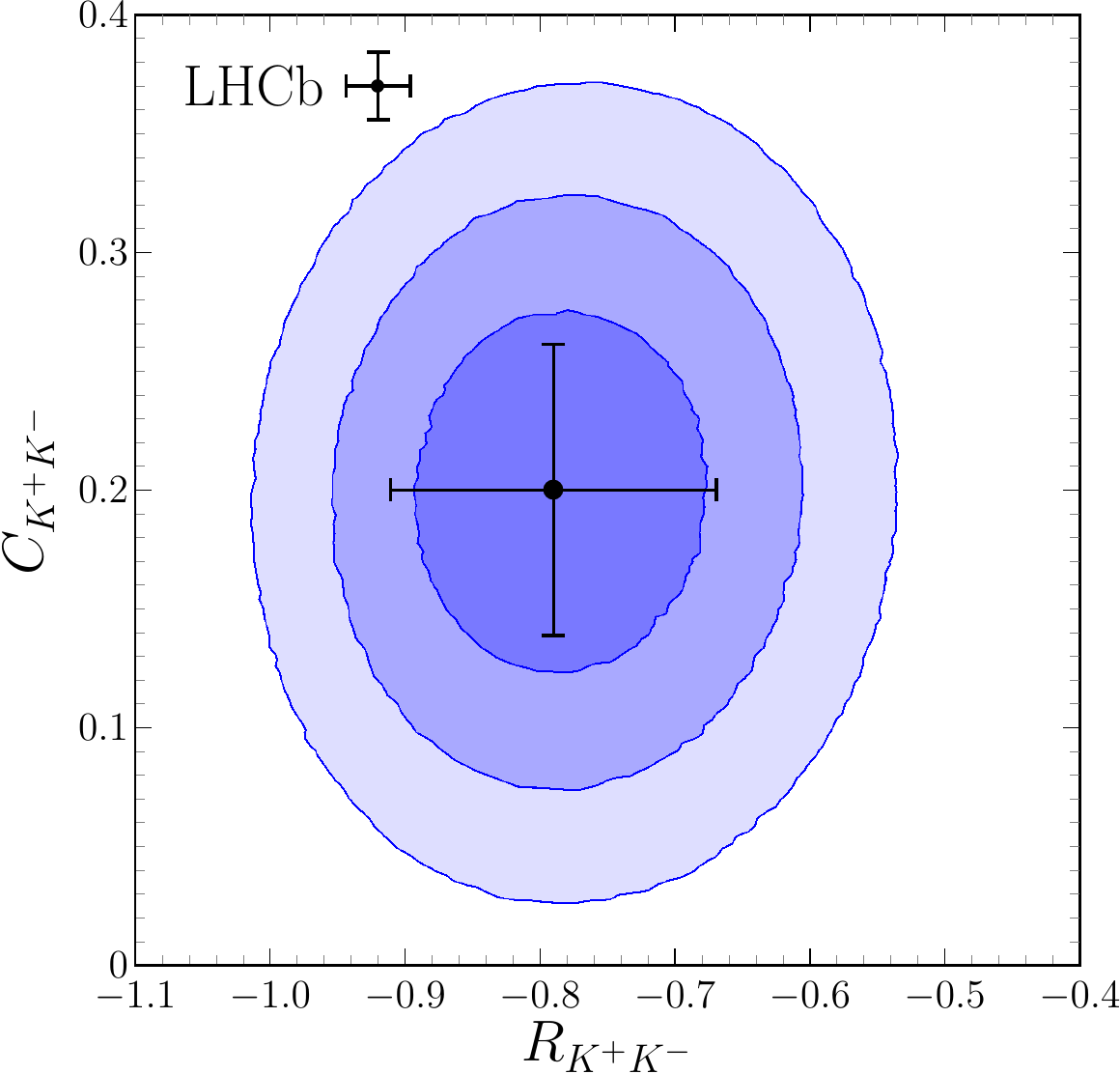}}\quad
\subfigure[$\SLHCb$ vs. $\CLHCb$.\label{sFIG:S:C}]{\includegraphics[height=0.28\textwidth]{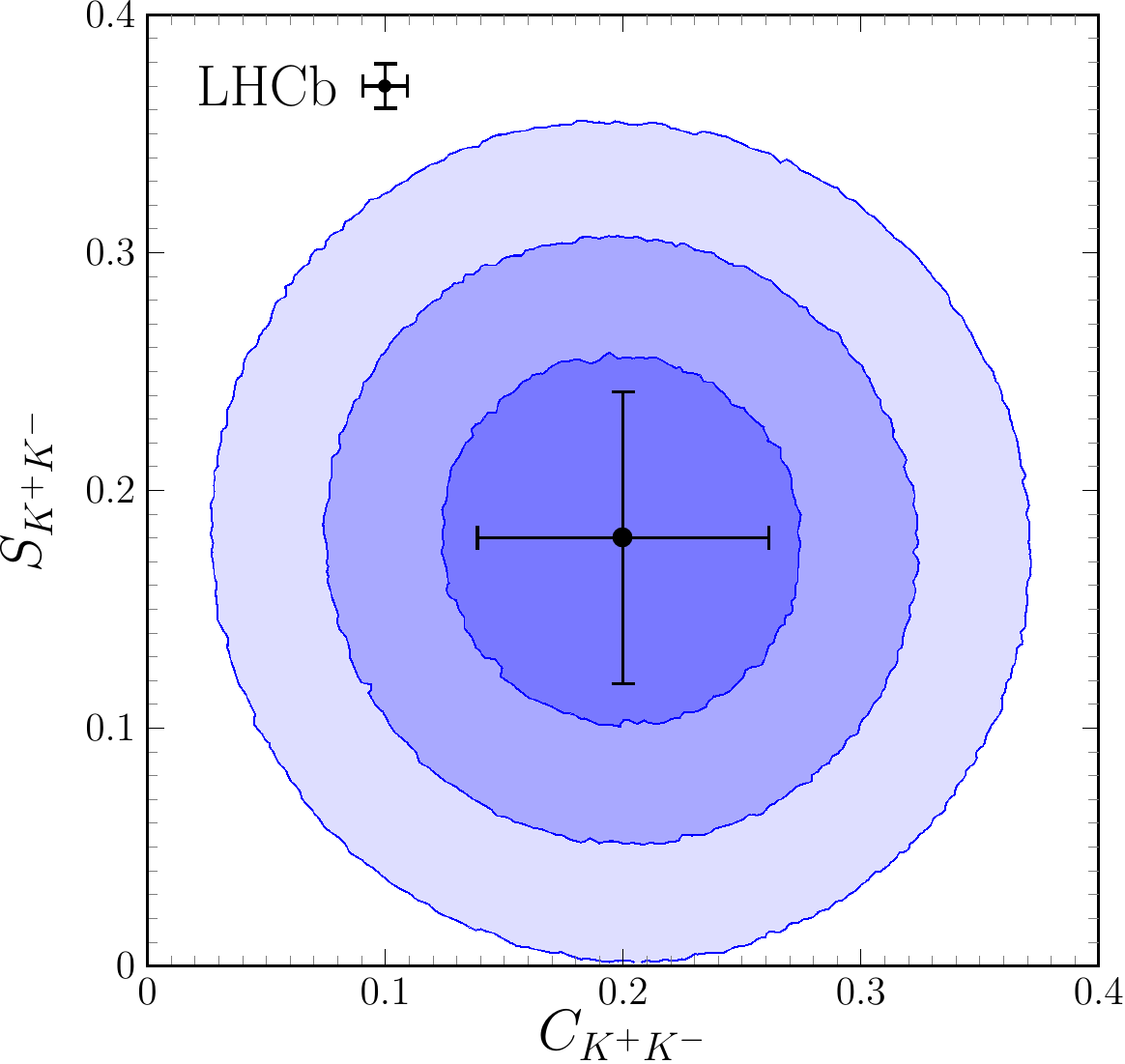}}
\caption{Validation: fitting back \refEQS{eq:rate:untagged}--\eqref{eq:rate:B} to the ``LHCb-like input data''; darker to lighter regions correspond to 1, 2 and 3 $\sigma$ for 2D-$\Delta\chi^2$.\label{FIG:fitback}}
\end{center}
\end{figure}

In order to perform analyses, some working input based on the available LHCb information in \cite{Aaij:2018tfw} is necessary. We adopt the following procedure to generate data reproducing the most relevant experimental aspects.
\begin{enumerate}
\item $\CLHCb$, $\SLHCb$ and $\RLHCb$ are set to the central values in \refEQ{eq::LHCb:pars:00}. 
\item A sample of 36840 events $B_s\to K^+K^-$, in accordance with\cite{Aaij:2018tfw}, is randomly generated; LHCb quotes a ``tagging power'' of 3.65 \% in \cite{Aaij:2018tfw}: consequently, $0.0365\times 36840\simeq 1345$ events are assigned to separate \emph{tagged} $\Bsb\to K^+K^-$ or $\Bs\to K^+K^-$ subsamples following, respectively, the time dependent rates
\begin{multline}\label{eq:rate:Bbar}
\Gamma(\Bsb(t)\to K^+K^-)\propto e^{-\GBs t}\left[\begin{matrix} \ChDGBs+\RLHCb\ShDGBs\\-\CLHCb\CcDMBs+\SLHCb\ScDMBs\end{matrix}\right],
\end{multline}
\begin{multline}\label{eq:rate:B}
\Gamma(\Bs(t)\to K^+K^-)\propto e^{-\GBs t}\left[\begin{matrix} \ChDGBs+\RLHCb\ShDGBs\\ +\CLHCb\CcDMBs-\SLHCb\ScDMBs\end{matrix}\right],
\end{multline}
The remaining $\simeq 35495$ events are assigned to the \emph{untagged} subsample, which follows the time dependent rate
\begin{multline}\label{eq:rate:untagged}
\Gamma(B_s(t)\to K^+K^-)_{\rm unt}=\Gamma(\Bsb(t)\to K^+K^-)+\Gamma(\Bs(t)\to K^+K^-)\\ \propto e^{-\GBs t}\left[\ChDGBs+\RLHCb\ShDGBs\right].
\end{multline}
For these untagged events, the flavour of the initial $B_s$ meson is not known. 
\item In order to incorporate the effect of the experimental time resolution, a smearing is applied: the value of $t$ associated to each event is randomly shifted following a gaussian distribution with mean $0$ and width $60$ fs. The considered range of times $t$ for these events is $[1;11]$ ps.
\item For each of the three subsamples, the corresponding distribution is binned (in bins of equal size, $100$ fs) giving $[\Gamma_{\rm unt}]_j$, $[\bar\Gamma]_j$ and $[\Gamma]_j$ with
\begin{align}
&\Gamma(B_s(t)\to K^+K^-)_{\rm unt}\mapsto [\Gamma_{\rm unt}]_j\,,\\ 
&\Gamma(\Bsb(t)\to K^+K^-)\mapsto [\bar\Gamma]_j\,,\\ 
&\Gamma(\Bs(t)\to K^+K^-)\mapsto [\Gamma]_j\,,
\end{align}
$j$ labels the 100 bins in the selected range $[1;11]$ ps.
\item A binned CP asymmetry $[A_{CP}]_j$ is computed
\begin{equation}\label{eq:LHCbdata:asymmetry}
[A_{CP}]_j=\frac{[\bar\Gamma]_j-[\Gamma]_j}{[\bar\Gamma]_j+[\Gamma]_j}\,.
\end{equation}
\item Iterating steps 2 to 5, we obtain distributions of $[\Gamma_{\rm unt}]_j$, $[\bar\Gamma]_j$, $[\Gamma]_j$ and $[A_{CP}]_j$: their means and standard deviations constitute the ``measurements'' with their statistical uncertainties, referred to, in the following, as the ``LHCb-like input data''.
\end{enumerate}
As in the LHCb analysis, $\DMBs=17.757$ ps$^{-1}$ and $\DGBs=0.083$ ps$^{-1}$ are used. 
For completeness, the resulting ``LHCb-like input data'' is shown in Appendix \ref{APP:dataLHCb}.
\newline For validation of the procedure, $\{\CLHCb,\SLHCb,\RLHCb\}$ are fitted back to the ``LHCb-like input data'' using the time dependences\footnote{The normalization of the untagged rate is a free parameter in the fit, that is, what one is really fitting is the time-dependent shape in \refEQ{eq:rate:untagged}. Then, the normalization of the tagged rates in \refEQS{eq:rate:Bbar}--\eqref{eq:rate:B} is not independent: it is fixed in terms of the normalization of the untagged rate (multiplying by the tagging power and dividing by 2).} in \refEQ{eq:rate:untagged} and \refEQS{eq:rate:Bbar}--\eqref{eq:rate:B}. The same smearing in the time dependence is considered (by convolving with the same gaussian distribution used in the generation of the ``LHCb-like input data'') and the same binning is applied. A $\chi^2$ function is constructed from the comparison of (i) the predicted binned rates and asymmetry, and (ii) the ``LHCb-like input data''.\\ 
 The results of this validation fit are shown in Figure \ref{FIG:fitback}, together with the LHCb input values. All three input parameters $\{\CLHCb,\SLHCb,\RLHCb\}$ are recovered with uncertainties in agreement with the LHCb result: the procedure simulates properly the LHCb sensitivity to the different time dependences.

%

\section{General Analysis with CPT Violation\label{SEC:analysis}}
\begin{figure}[h!tb]
\begin{center}
\subfigure[$\Lkk$.\label{sFIG:d0:imLf:reLf}]{\includegraphics[height=0.33\textwidth]{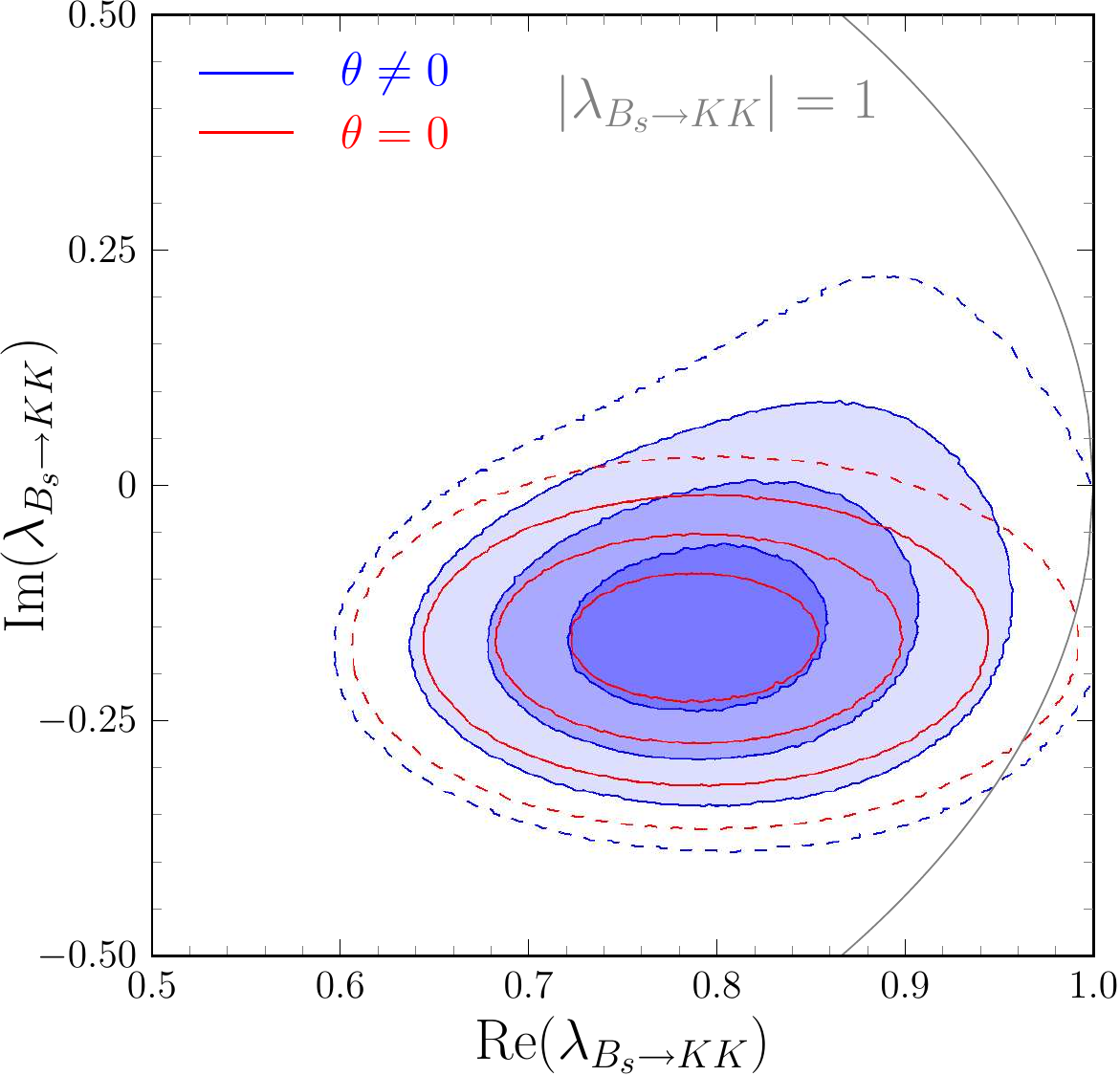}}\qquad
\subfigure[$\im{\thBs}$ vs. $\re{\thBs}$.\label{sFIG:d0:imth:reth}]{\includegraphics[height=0.33\textwidth]{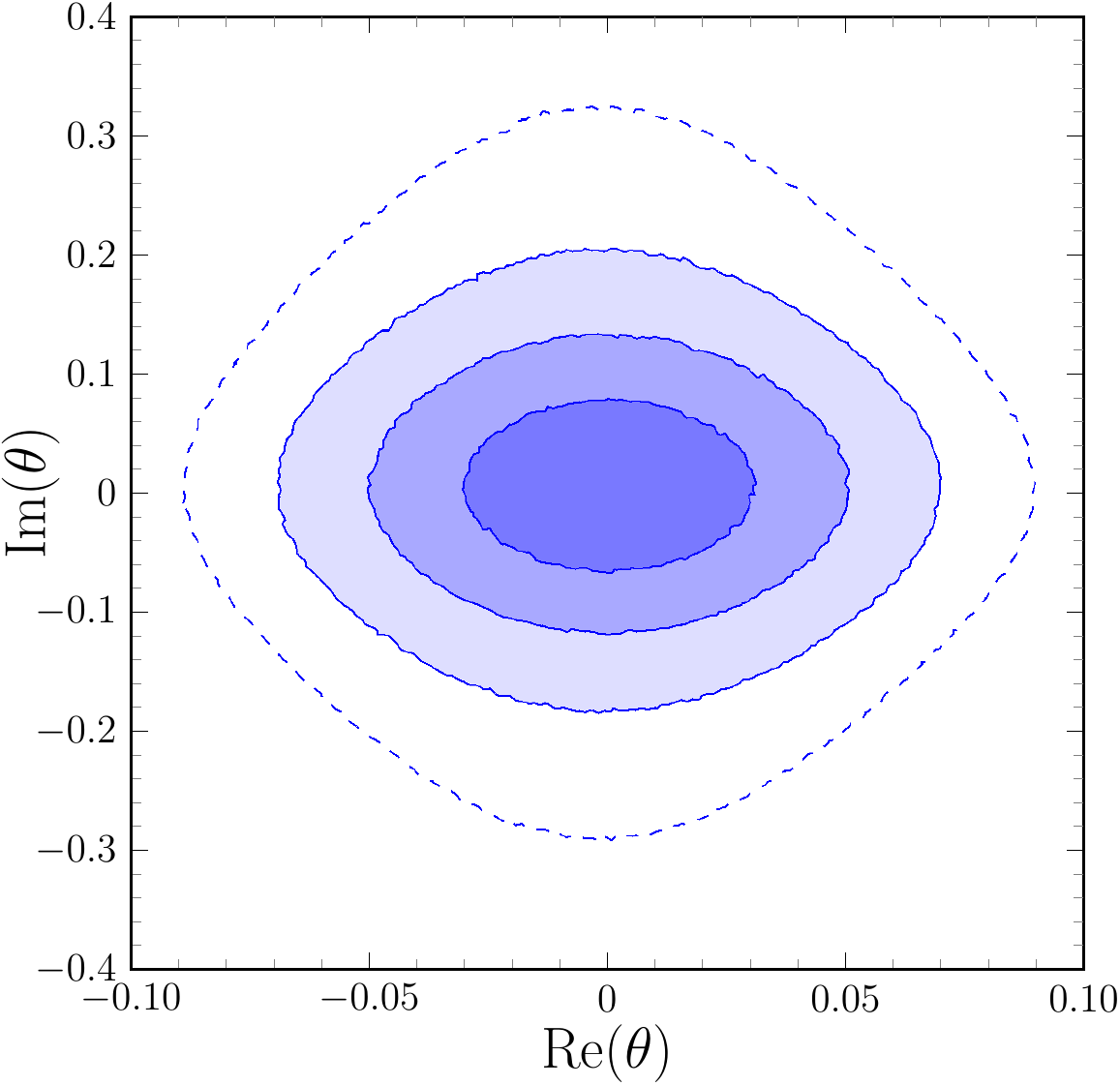}}
\caption{Fit results (I); in blue, darker to lighter regions correspond to 1, 2 and 3 $\sigma$ for 2D-$\Delta\chi^2$, the dashed contours correspond to 4 $\sigma$. In red, the corresponding contours for fixed $\thBs=0$ are shown.\label{FIG:res:d0:I}}
\end{center}
\end{figure}

\begin{figure}[h!tb]
\begin{center}
\subfigure[$\im{\thBs}$ vs. $\Ckk$.\label{sFIG:d0:imth:CKK}]{\includegraphics[height=0.28\textwidth]{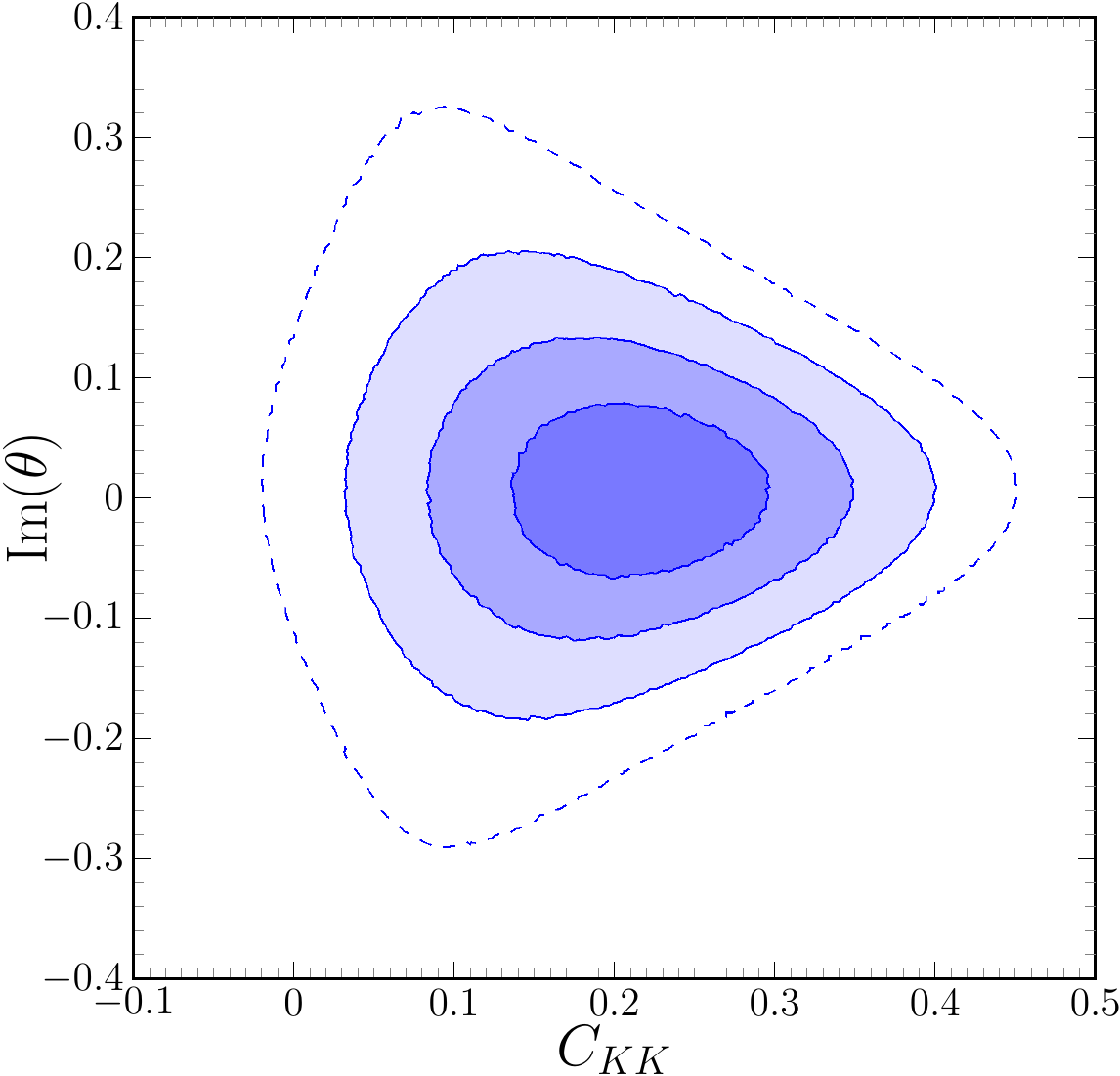}}\quad
\subfigure[$\im{\thBs}$ vs. $\Skk$.\label{sFIG:d0:imth:SKK}]{\includegraphics[height=0.28\textwidth]{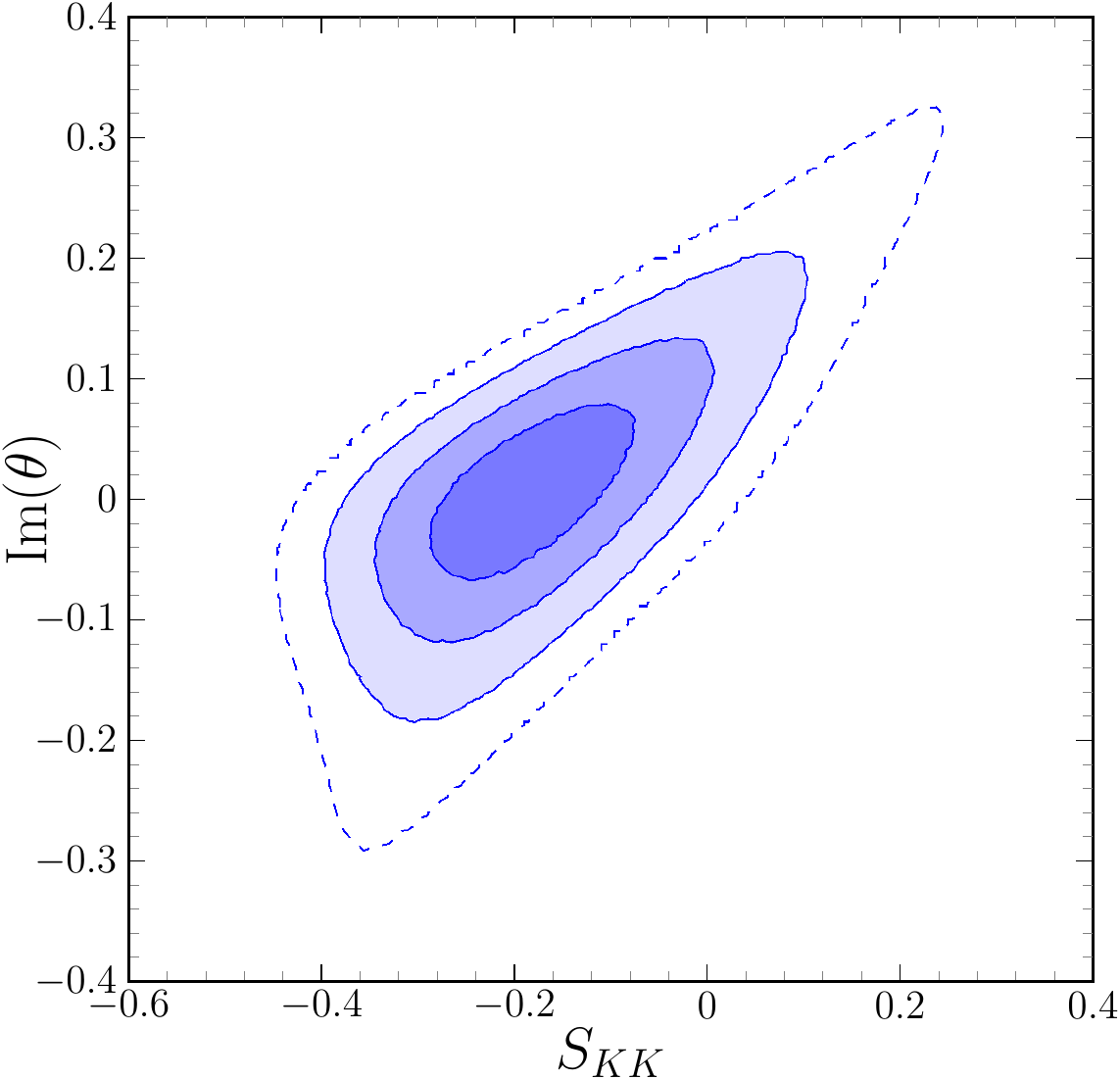}}\quad
\subfigure[$\im{\thBs}$ vs. $\Rkk$.\label{sFIG:d0:imth:RKK}]{\includegraphics[height=0.28\textwidth]{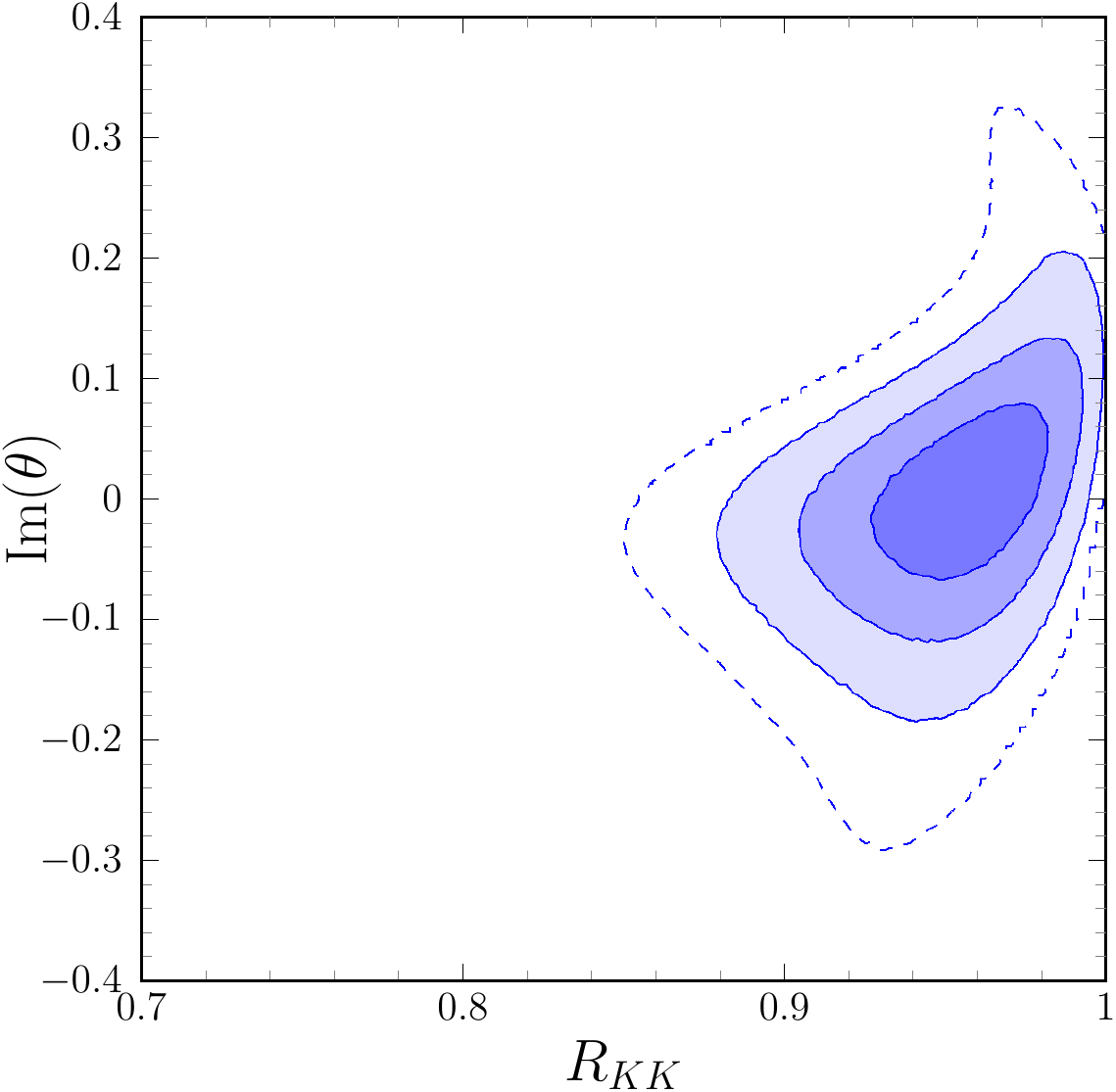}}
\caption{Fit results (II) (conventions as in Figure \ref{FIG:res:d0:I}).\label{FIG:res:d0:II}}
\end{center}
\end{figure}

\begin{figure}[h!tb]
\begin{center}
\subfigure[$\Skk$ vs. $\Rkk$.\label{sFIG:d0:SKK:RKK}]{\includegraphics[height=0.28\textwidth]{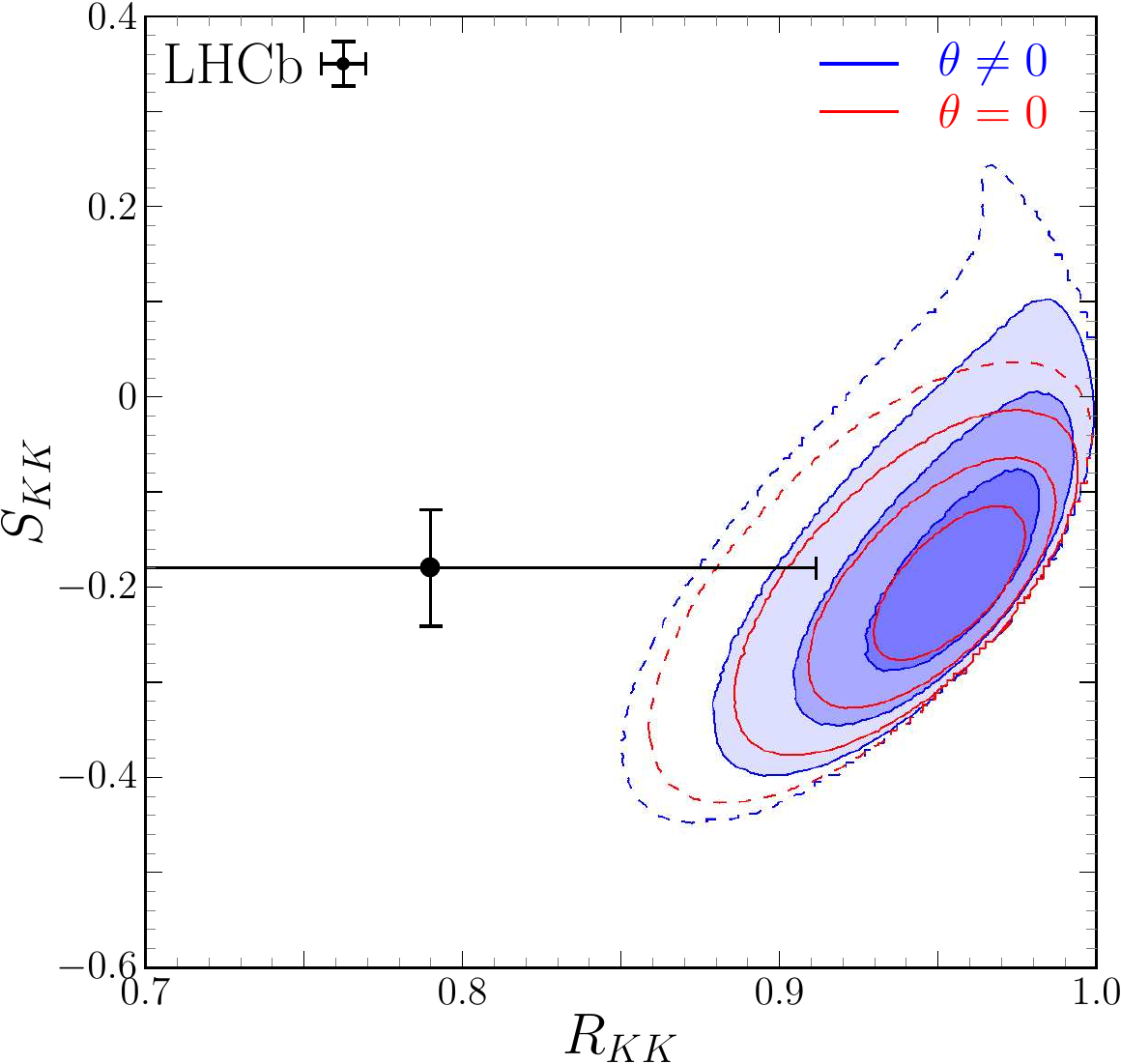}}\quad
\subfigure[$\Ckk$ vs. $\Rkk$.\label{sFIG:d0:CKK:RKK}]{\includegraphics[height=0.28\textwidth]{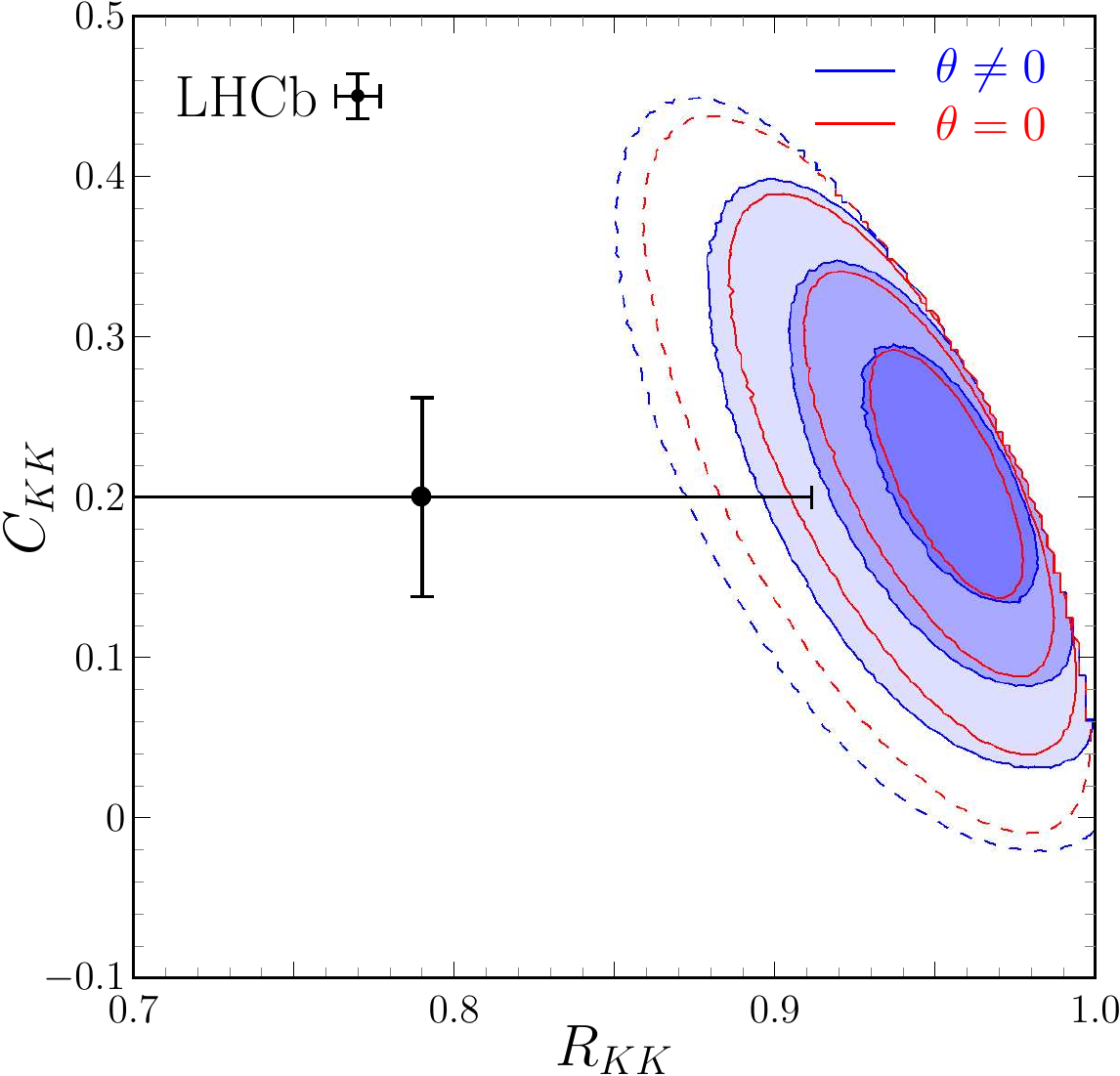}}\quad
\subfigure[$\Skk$ vs. $\Ckk$.\label{sFIG:d0:SKK:CKK}]{\includegraphics[height=0.28\textwidth]{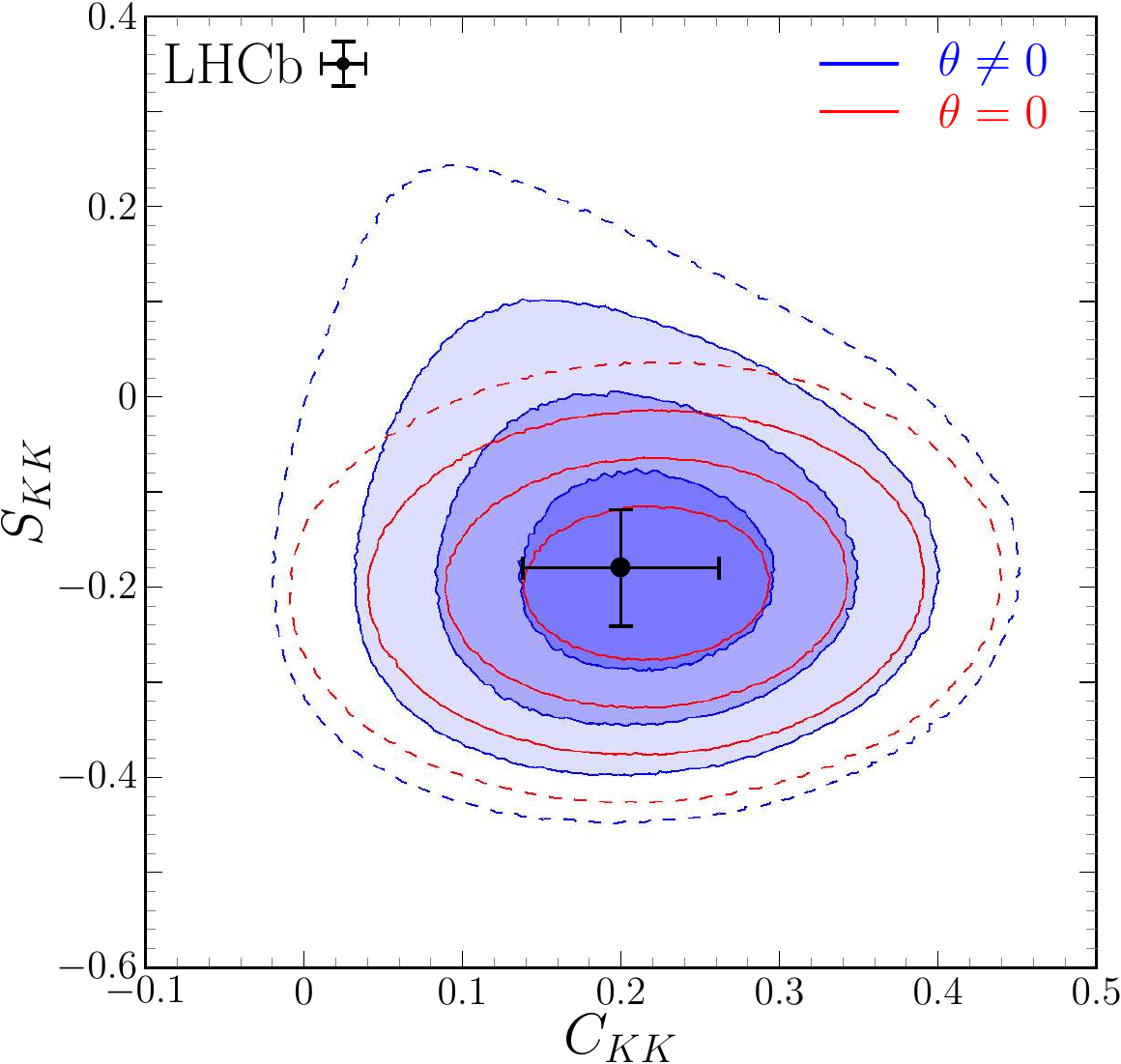}}
\caption{Fit results (III) (conventions as in Figure \ref{FIG:res:d0:I}).\label{FIG:res:d0:III}}
\end{center}
\end{figure}

With the ``LHCb-like input data'' discussed in the previous section, we can now perform fits using the general expressions in section \ref{SEC:gen}. These expressions depend on $\Lkk$, $\dBs$ and $\thBs$: the analysis assumes $\dBs=0$, as done by LHCb in \cite{Aaij:2018tfw}; in appendix \ref{APP:analysis:d} the effect of relaxing this assumption is explored. The predictions computed in terms of $\Lkk$, $\dBs= 0$ and $\thBs$ are smeared and binned in time, exactly as in the generation of the input data of the previous section. A $\chi^2$ function is then constructed from the comparison of the computed rates and asymmetry and the ``LHCb-like input data''. In the fit, the minimum of the $\chi^2$ function together with regions around it corresponding to different values of $\Delta\chi^2=\chi^2-\chi^2_{\rm Min}$ are obtained.\\ 
The results of the fit (with $\dBs=0$) are shown in figures \ref{FIG:res:d0:I}, \ref{FIG:res:d0:II} and \ref{FIG:res:d0:III}; results for $\thBs=0$ are also displayed for comparison.\\
In figure \ref{FIG:res:d0:I}, the allowed regions for $\Lkk$ and $\thBs$ are shown. It is to be noticed that:
\begin{itemize}
\item the real part of $\thBs$ is quite constrained, bound at the $10^{-2}$ level, while for the imaginary part larger values are still allowed;
\item concerning $\Lkk$, it is clear that with $\thBs\neq 0$ the phase of $\Lkk$ is less constrained, while $\abs{\Lkk}$ is not much affected. It is also important to notice that $\abs{\Lkk}=1$ is excluded at a level between $3\sigma$ and $4\sigma$, and that the CP conserving scenario with $(\re{\Lkk},\im{\Lkk})=(1,0)$ is excluded at around the $4 \sigma$ level for the general case $\thBs\neq 0$; for $\thBs=0$ the significance of that exclusion rises to the $\sim 5\sigma$ level.
\end{itemize}
Figure \ref{FIG:res:d0:II} shows the allowed regions for $\im{\thBs}$ vs. $\Ckk$, $\Skk$, $\Rkk$ (defined as in \refEQ{eq:CfSfRf:00} in terms of $\Lkk$). While for $\Ckk$, and thus $\abs{\Lkk}$, the only noticeable correlation is that for large $\Ckk$ values, the allowed range for $\im{\thBs}$ is reduced, it is clear that the obtained $\Skk$ and $\Rkk$, and thus $\arg(\Lkk)$, are on the contrary quite correlated with $\im{\thBs}$. Attending to \refEQ{eq:DIF:lin:02}, the sensitivity to $\ScDMBs$ constrains the combination $\Skk-\im{\thBs}$, rather than $\Skk$ and $\im{\thBs}$ separately, as can be seen in figure \ref{sFIG:d0:imth:SKK}.

\noindent Figure \ref{FIG:res:d0:III} shows the allowed regions for the different pairs in $\{\Ckk,\Skk,\Rkk\}$, further illustrating the previous comments.

\noindent From these results, the most relevant aspects are the following.
\begin{itemize}
\item For CPT violation, significant bounds can be obtained on $\thBs$: $\re{\thBs}=(0.0\pm 2.0)\times 10^{-2}$ and $\im{\thBs}=\left(0.0\begin{smallmatrix}+5.0\\-4.0\end{smallmatrix}\right)\times 10^{-2}$; bounds at the $10^{-2}$ level were also obtained in the analysis of \cite{Aaij:2016mos} for the parameter $\thBs$, although in a different theoretical context where $\re{\thBs}$ and $\im{\thBs}$ are not independent.
\item Both $\Ckk$ and $\Skk$ correspond rather closely to the LHCb parameters $\CLHCb$ and $\SLHCb$, while that is not the case for $\Rkk$ and $\RLHCb$.
\item With $\thBs\neq 0$, the uncertainty in $\Skk$ (i.e. in the phase of the mixing $\times$ decay amplitude, $\arg(\Lkk)$), is larger than that of the LHCb parameter $\SLHCb$: indirect CP violation does only appear at the $2\sigma$ level.
\item Nevertheless, it is clear that, regardless of CPT violation, $\Ckk\neq 0$, i.e. $\abs{\Lkk}\neq 1$, establishing the observation of direct CP violation in $B_s\to K^+K^-$ close to the $4\sigma$ level.
\end{itemize}

%
%

\section*{Conclusions\label{SEC:conc}}
We have presented a simple analysis based on recent LHCb work \cite{Aaij:2018tfw}, pointing out the benefits of adopting a fully consistent theoretical description of the \BBsmix\ mixing. Among the consequences of doing so, one should stress that direct CP violation in $B_s\to K^+K^-$ could be robustly established and that bounds on CPT violation in the \BBsmix\ system can be obtained. We suggest our experimental colleagues from the LHCb collaboration to analyse $B_s\to K^+K^-$ transitions in terms of the complete time dependences discussed in section \ref{SEC:gen}, including potential CPT violating effects.

\section*{Acknowledgments}
The authors thank F. Martínez Vidal for extensive discussions and critical reading of the manuscript, and Clara Remón for fruitful discussions.\\ 
This work is partially supported by Spanish MINECO under grant FPA2015-68318-R, FPA2017-85140-C3-3-P and by the Severo Ochoa Excellence Center Project SEV-2014-0398, 
 and by Funda\c{c}\~ao para a Ci\^encia e a Tecnologia (FCT, Portugal) through the projects UID/FIS/00777/2019, CERN/FIS-PAR/0004/2017 and PTDC/FIS-PAR/29436/2017 which are partially funded through POCTI (FEDER), COMPETE, QREN and EU. 
MN acknowledges support from FCT through postdoctoral grant SFRH/BPD/112999/2015.

\appendix

\clearpage
\section{General expressions\label{APP:gen}}
The coefficients in \refEQS{eq:evol:NM:00}-\eqref{eq:evol:NMb:00} read
\begin{align}
\label{eq:Ch:gen:00}
&\COCh{\pm}{f}=\frac{1\pm\delta}{2(1-\Cf\delta)}
\left\{\begin{matrix}
(1\mp\Cf)(1+\abs{\theta}^2)+(1\pm\Cf)\abs{1-\theta^2}\\ \pm 2\Rf\re{\theta^\ast\sqrt{1-\theta^2}}+2\Sf\im{\theta^\ast\sqrt{1-\theta^2}}
\end{matrix}\right\},\\
\label{eq:Cc:gen:00}
&\COCc{\pm}{f}=\frac{1\pm\delta}{2(1-\Cf\delta)}
\left\{\begin{matrix}
(1\mp\Cf)(1-\abs{\theta}^2)-(1\pm\Cf)\abs{1-\theta^2}\\ \mp 2\Rf\re{\theta^\ast\sqrt{1-\theta^2}}-2\Sf\im{\theta^\ast\sqrt{1-\theta^2}}
\end{matrix}\right\},\\
\label{eq:Sh:gen:00}
&\COSh{\pm}{f}=\frac{1\pm\delta}{(1-\Cf\delta)}
\left\{\begin{matrix}
(\Cf\mp 1)\re{\theta}\\ -\Rf\re{\sqrt{1-\theta^2}}\mp \Sf\im{\sqrt{1-\theta^2}}
\end{matrix}\right\},\\
\label{eq:Sc:gen:00}
&\COSc{\pm}{f}=\frac{1\pm\delta}{(1-\Cf\delta)}
\left\{\begin{matrix}
-(\Cf\mp 1)\im{\theta}\\ \mp \Sf\re{\sqrt{1-\theta^2}}+\Rf\im{\sqrt{1-\theta^2}}
\end{matrix}\right\},
\end{align}
with $\COCh{+}{f}=\COCh{\NMb}{f}$, $\COCh{-}{f}=\COCh{\NM}{f}$, and similarly for $\COChBASE\to\COShBASE,\COCcBASE,\COScBASE$.\\
Following \refEQS{eq:Ch:gen:00}--\eqref{eq:Sc:gen:00}, the general expressions for the coefficients in $\ACP{M\to f}(t)$ are
\begin{align}
&\DIFCh=
\left\{\begin{matrix}
-\Cf(1+\abs{\theta}^2-\abs{1-\theta^2})+2\Rf\re{\theta^\ast\sqrt{1-\theta^2}}\\ +\delta[1+\abs{\theta}^2+\abs{1-\theta^2}+2\Sf\im{\theta^\ast\sqrt{1-\theta^2}}]
\end{matrix}\right\},\\
&\DIFCc=
\left\{\begin{matrix}
-\Cf(1-\abs{\theta}^2+\abs{1-\theta^2})-2\Rf\re{\theta^\ast\sqrt{1-\theta^2}}\\ +\delta[1-\abs{\theta}^2-\abs{1-\theta^2}-2\Sf\im{\theta^\ast\sqrt{1-\theta^2}}]
\end{matrix}\right\},\\
&\DIFSh=2
\left\{\begin{matrix}
-\re{\theta}-\Sf\im{\sqrt{1-\theta^2}}\\ +\delta[\Cf\re{\theta}-\Rf\re{\sqrt{1-\theta^2}}]
\end{matrix}\right\},\\
&\DIFSc=2
\left\{\begin{matrix}
\im{\theta}-\Sf\re{\sqrt{1-\theta^2}}\\ +\delta[-\Cf\im{\theta}+\Rf\im{\sqrt{1-\theta^2}}]
\end{matrix}\right\},
\end{align}
and
\begin{align}
&\SUMCh=
\left\{\begin{matrix}
1+\abs{\theta}^2+\abs{1-\theta^2}+2\Sf\im{\theta^\ast\sqrt{1-\theta^2}}\\ +\delta[-\Cf(1+\abs{\theta}^2-\abs{1-\theta^2})+2\Rf\re{\theta^\ast\sqrt{1-\theta^2}}]
\end{matrix}\right\},\\
&\SUMCc=
\left\{\begin{matrix}
1-\abs{\theta}^2-\abs{1-\theta^2}-2\Sf\im{\theta^\ast\sqrt{1-\theta^2}}\\ +\delta[-\Cf(1-\abs{\theta}^2+\abs{1-\theta^2})-2\Rf\re{\theta^\ast\sqrt{1-\theta^2}}]
\end{matrix}\right\},\\
&\SUMSh=2
\left\{\begin{matrix}
\Cf\re{\theta}-\Rf\re{\sqrt{1-\theta^2}}\\ -\delta[\re{\theta}+\Sf\im{\sqrt{1-\theta^2}}]
\end{matrix}\right\},\\
&\SUMSc=2
\left\{\begin{matrix}
-\Cf\im{\theta}+\Rf\im{\sqrt{1-\theta^2}}\\ +\delta[\im{\theta}-\Sf\re{\sqrt{1-\theta^2}}]
\end{matrix}\right\}.
\end{align}

%

\clearpage
\section{``LHCb-like input data''\label{APP:dataLHCb}}
\begin{figure}[htb!]
\begin{center}
\subfigure[Untagged $\Bsb,\Bs\to K^+K^-$ rate.\label{FIG:LHCb:data:BsKK:untagged}]{\includegraphics[width=0.6\textwidth]{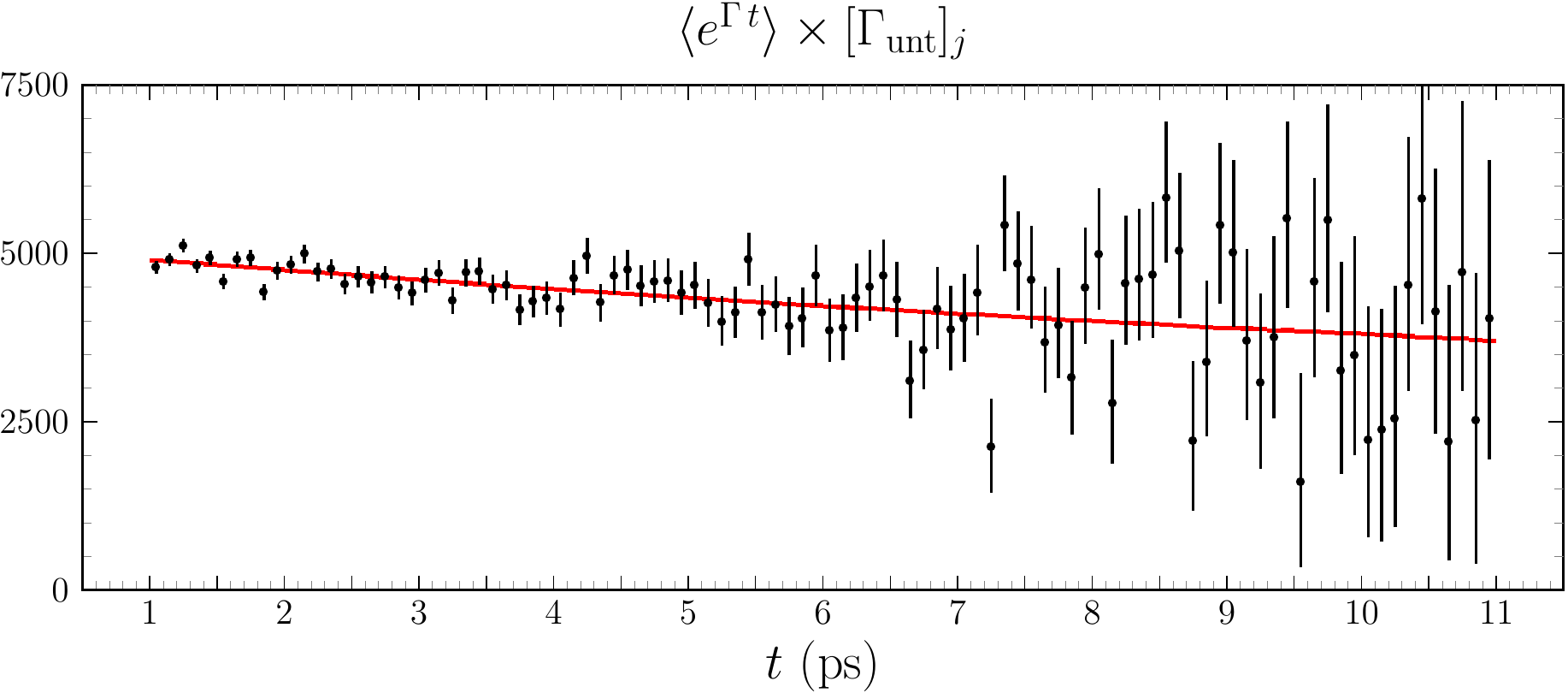}}\\
\begin{minipage}{0.52\textwidth}
\subfigure[Tagged $\Bsb\to K^+K^-$ rate.\label{FIG:LHCb:data:BbarsKK:tagged}]{\includegraphics[width=0.95\textwidth]{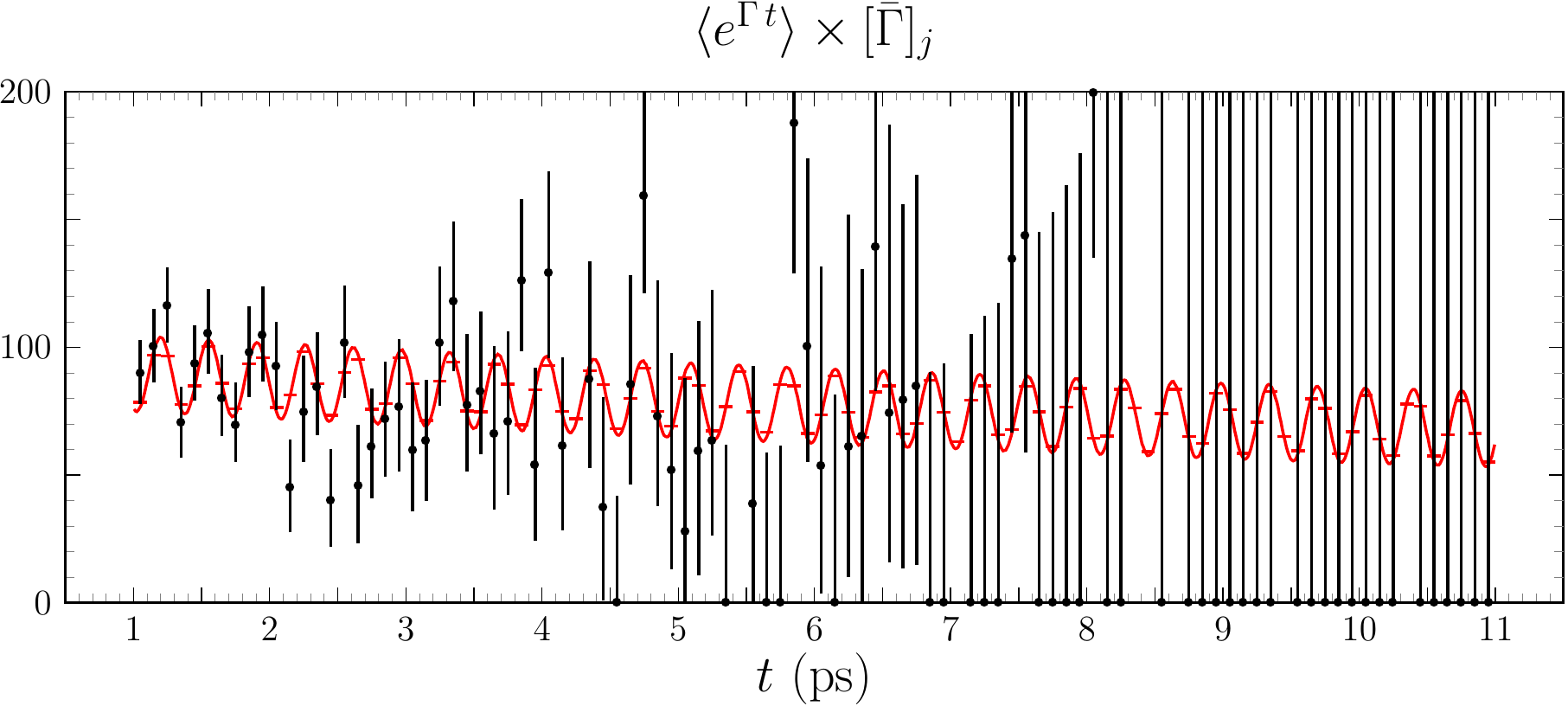}}\\
\subfigure[Tagged $\Bs\to K^+K^-$ rate.\label{FIG:LHCb:data:BsKK:tagged}]{\includegraphics[width=0.95\textwidth]{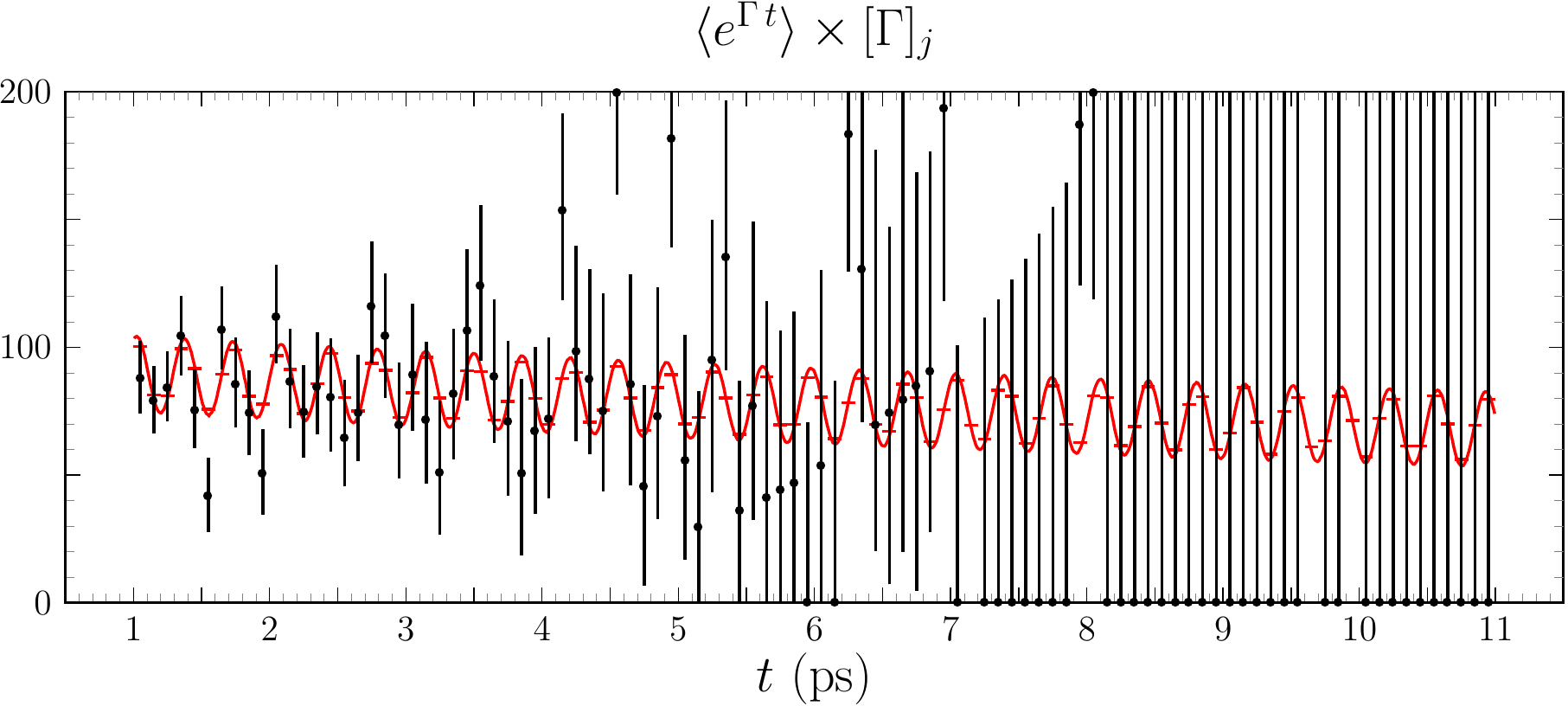}}
\end{minipage}
\begin{minipage}{0.45\textwidth}
\subfigure[CP asymmetry $\ACP{B_s\to KK}$.\label{FIG:LHCb:data:BsKK:asymmetry}]{\includegraphics[width=0.95\textwidth]{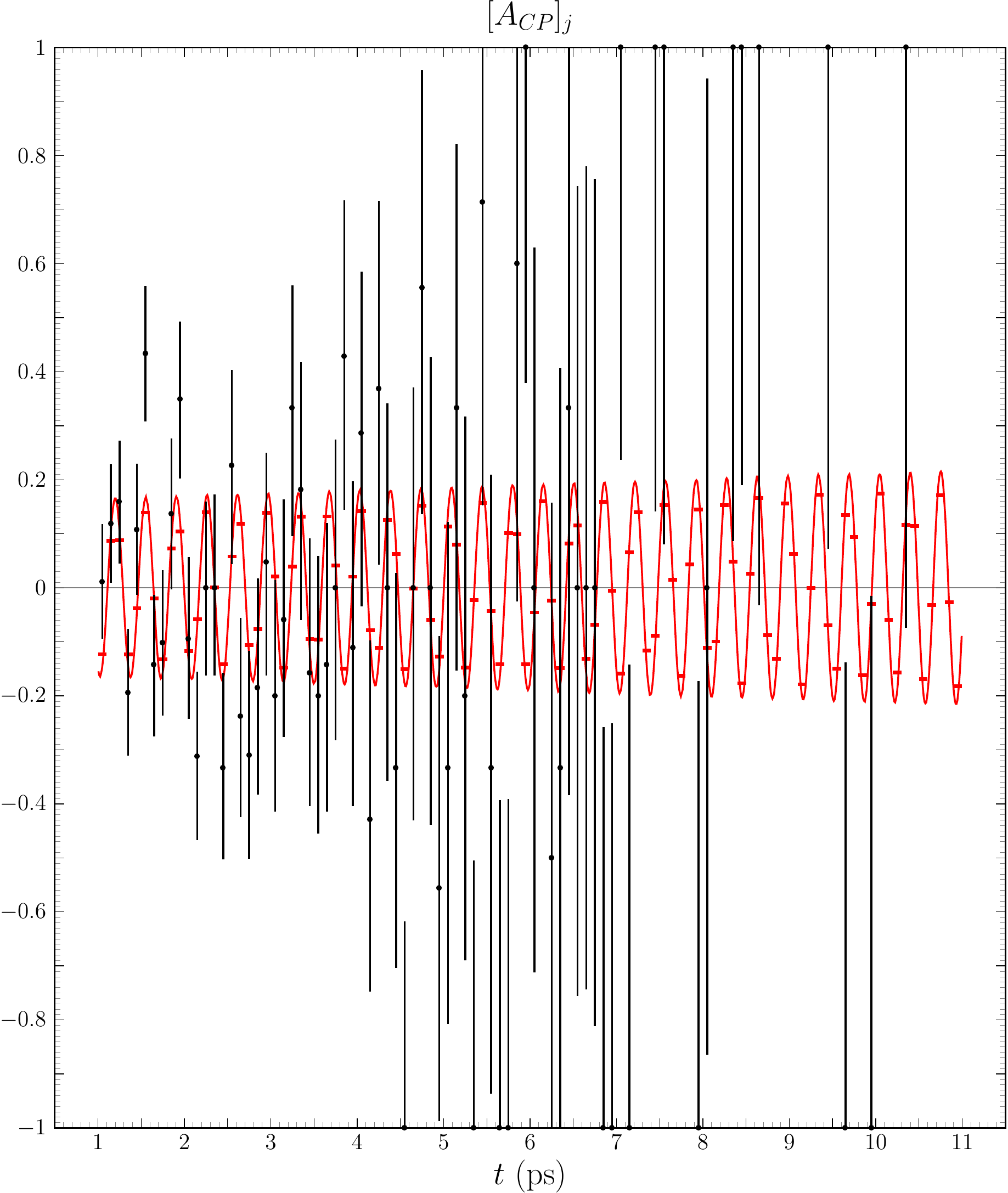}}
\end{minipage}
\caption{``LHCb-like input data'' for $\Bs,\Bsb\to K^+K^-$: the values and the errors of an example sample are represented with, respectively, black dots and vertical bars; the horizontal red bars show the mean value over the binned generated samples and the red curve the corresponding computed expectation. In \ref{FIG:LHCb:data:BsKK:untagged} to \ref{FIG:LHCb:data:BsKK:tagged} the values of the observables are represented multiplied by the average of $\exp(\GBs t)$ in the corresponding bin for improved readability. One can observe the mixing induced oscillations with period $2\pi/\DMBs\sim 0.35$ ps (28 oscillations in the range $[1;11]$ ps) in Figures \ref{FIG:LHCb:data:BbarsKK:tagged} to \ref{FIG:LHCb:data:BsKK:asymmetry}.\label{FIG:LHCb:data}}
\end{center}
\end{figure}

Figure \ref{FIG:LHCb:data} illustrates the ``LHCb-like input data'' obtained with the procedure of section \ref{SEC:data}. Figure \ref{FIG:LHCb:data:BsKK:untagged} corresponds to the untagged rate in \refEQ{eq:rate:untagged}, while Figures \ref{FIG:LHCb:data:BbarsKK:tagged} and \ref{FIG:LHCb:data:BsKK:tagged} correspond to the tagged rates in \refEQS{eq:rate:Bbar}-\eqref{eq:rate:B}. For the CP asymmetry in \refEQ{eq:LHCbdata:asymmetry}, the result is shown in Figure \ref{FIG:LHCb:data:BsKK:asymmetry}. 

\clearpage
\section{General Analysis with CPT Violation and $\dBs\neq 0$ \label{APP:analysis:d}}
\begin{figure}[h!tb]
\begin{center}
\subfigure[$\Lkk$.\label{sFIG:d:imLf:reLf}]{\includegraphics[height=0.33\textwidth]{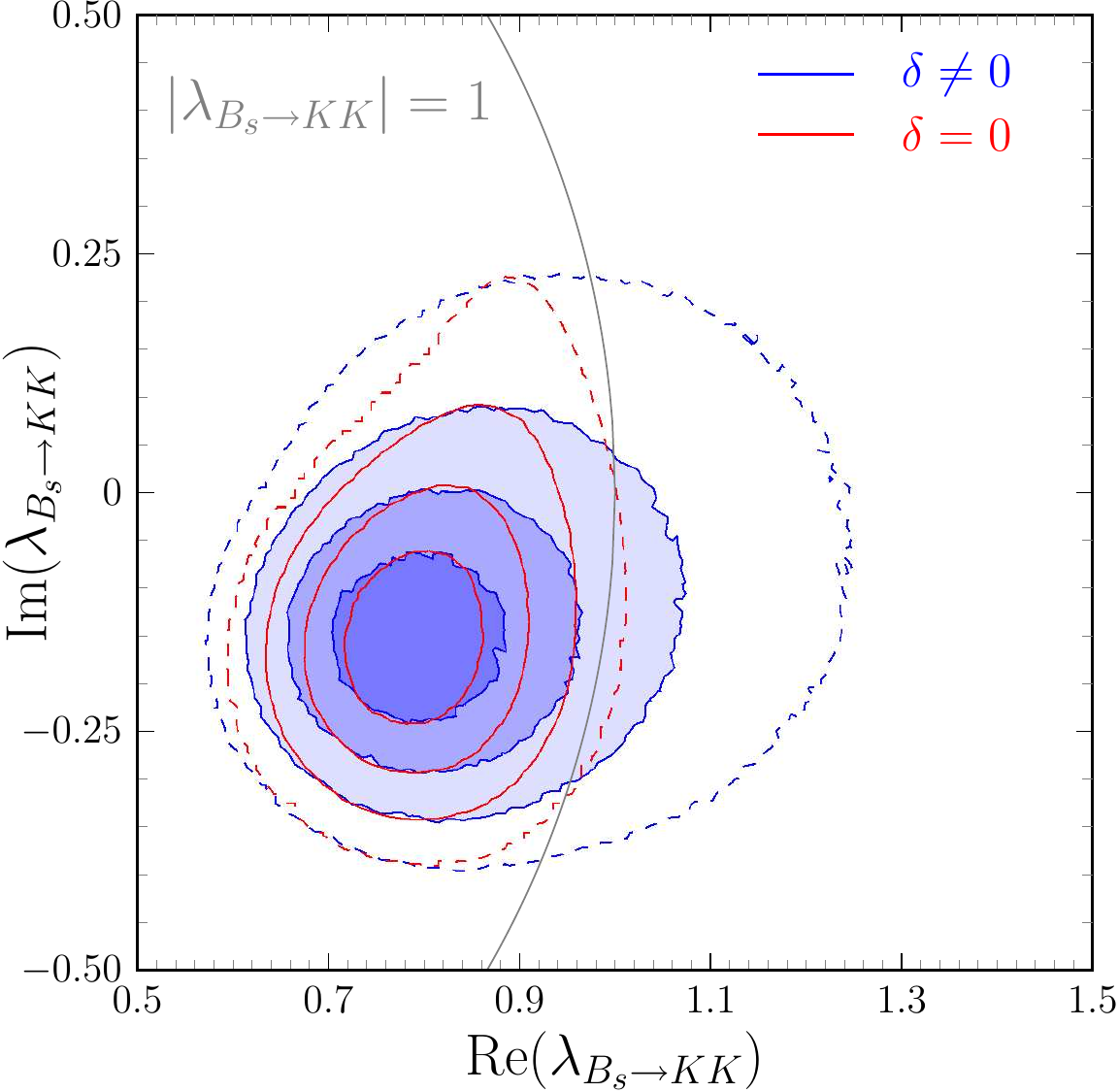}}\qquad
\subfigure[$\im{\thBs}$ vs. $\re{\thBs}$.\label{sFIG:d:imth:reth}]{\includegraphics[height=0.33\textwidth]{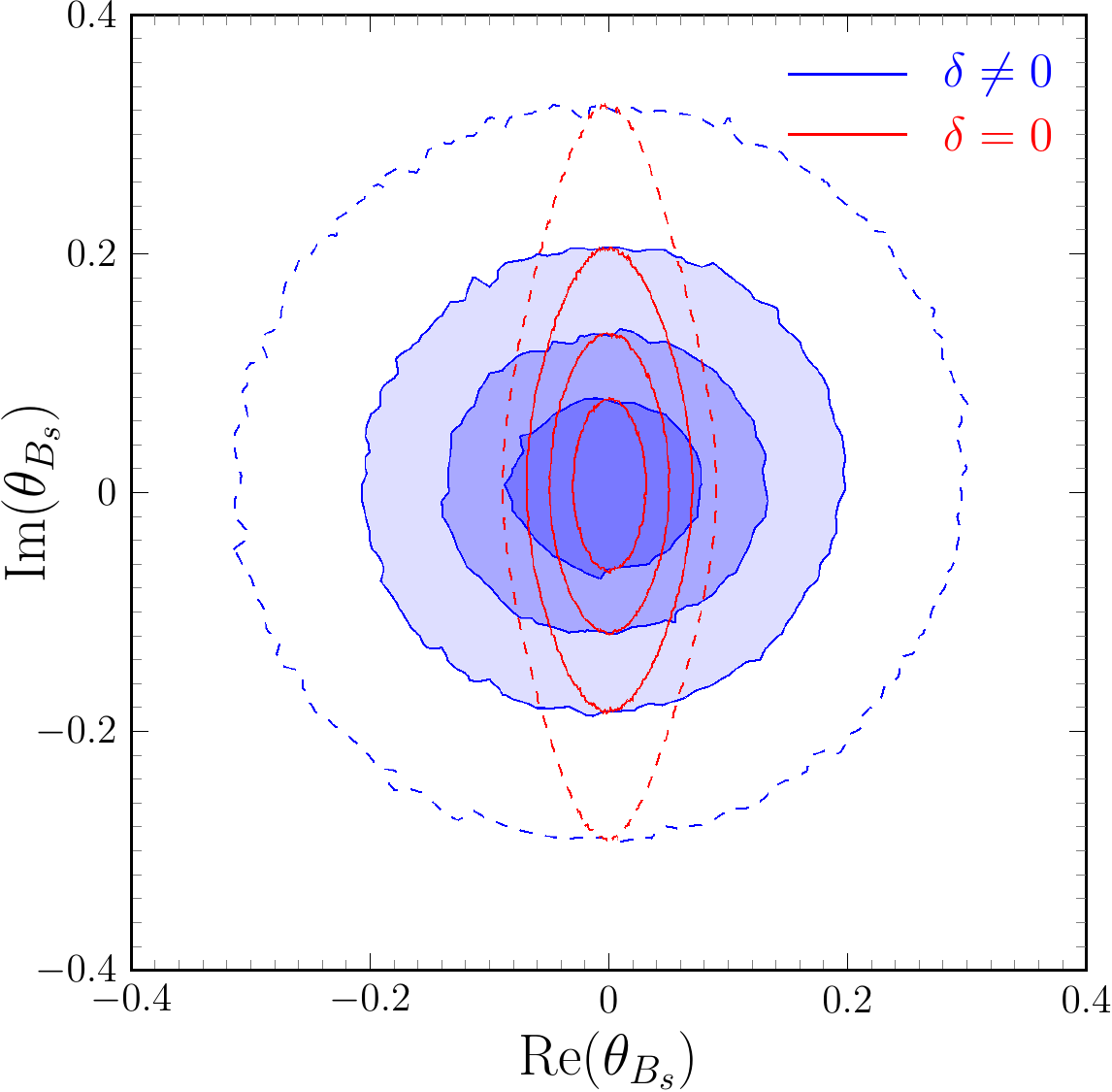}}
\caption{Fit results with $\dBs\neq 0$; in blue, darker to lighter regions correspond to 1, 2 and 3 $\sigma$ for 2D-$\Delta\chi^2$, the dashed contour corresponds to 4 $\sigma$. In red, the corresponding contours for the $\dBs=0$ analysis of section \ref{SEC:analysis} are shown.\label{FIG:res:d:I}}
\end{center}
\end{figure}

The analysis of section \ref{SEC:analysis} assumes, as the LHCb analysis in  \cite{Aaij:2018tfw}, $\dBs=0$. Constraints on $\dBs$ below the $10^{-2}$ level are obtained from semileptonic decays \cite{Aaij:2016yze}: they are obtained, however, under the assumption of $\thBs=0$, and thus they cannot be imposed here. In this appendix we illustrate the result of relaxing the assumption $\dBs=0$: the fit to the ``LHCb-like input data'' maintains $\dBs$ (in addition to $\Lkk$ and $\thBs$) in the general expressions of appendix \ref{APP:gen}. In terms of $\dBs$, equations \eqref{eq:Ch:gen:00}--\eqref{eq:Sc:gen:00} have the following common structure:
\begin{equation}
\COCh{\pm}{KK}=\frac{1\pm\dBs}{1-\Ckk\dBs}\left[\COCh{\pm}{KK}\right]_{\dBs=0},\quad\text{and similarly for}\ \COChBASE\to\COShBASE,\COCcBASE,\COScBASE,
\end{equation}
that is, all the $\dBs$ dependence is factored in $\frac{1\pm\dBs}{1-\Ckk\dBs}$ while $\left[\COCh{\pm}{KK}\right]_{\dBs=0}$,\ldots, are precisely the coefficients of the different time dependencies which enter the analysis of the previous subsection, with $\dBs=0$.
For the sums and differences of coefficients in \refEQ{eq:coefs:sum:dif:00} it follows immediately that
\begin{align}
&\SUMCh=\frac{1}{1-\Ckk\dBs}\left\{\left[\SUMCh\right]_{\dBs=0}+\dBs\,\left[\DIFCh\right]_{\dBs=0}\right\},\\
&\DIFCh=\frac{1}{1-\Ckk\dBs}\left\{\left[\DIFCh\right]_{\dBs=0}+\dBs\,\left[\SUMCh\right]_{\dBs=0}\right\},
\end{align}
and, of course, similarly for $\COChBASE\to\COShBASE,\COCcBASE,\COScBASE$. At linear order in $\thBs$ (except for $\dBs\thBs$ terms which are dropped), \refEQS{eq:DIF:lin:02}--\eqref{eq:SUM:lin:02} for $\dBs\neq 0$ read
\begin{alignat}{3}
\nonumber
&\DIFCh=2\frac{\Rkk\re{\thBs}+\dBs}{1-\Ckk\dBs},\qquad &&\DIFCc=-2\frac{\Ckk+\Rkk\re{\thBs}}{1-\Ckk\dBs},\\
\label{eq:DIFdelta:lin:02}
&\DIFSh=-2\frac{\re{\thBs}+\Rkk\dBs}{1-\Ckk\dBs},\qquad &&\DIFSc=2\frac{\im{\thBs}-\Skk}{1-\Ckk\dBs},
\end{alignat}
\begin{alignat}{3}
\nonumber
&\SUMCh=2\frac{1-\Skk\im{\thBs}}{1-\Ckk\dBs},\qquad &&\SUMCc=2\frac{\Skk\im{\thBs}-\Ckk\dBs}{1-\Ckk\dBs},\\
\label{eq:SUMdelta:lin:02}
&\SUMSh=2\frac{\Ckk\re{\thBs}-\Rkk}{1-\Ckk\dBs},\qquad &&\SUMSc=-2\frac{\Ckk\im{\thBs}+\Skk\dBs}{1-\Ckk\dBs}.
\end{alignat}
Summarising the effects of $\dBs\neq 0$: 
\begin{itemize}
\item The tagged rates are rescaled differently,
\begin{equation}
\Gamma(\Bsb(t)\to K^+K^-)=\frac{1+\dBs}{1-\Ckk\dBs}\left[\Gamma(\Bsb(t)\to K^+K^-)\right]_{\dBs=0}\,,
\end{equation}
\begin{equation}
\Gamma(\Bs(t)\to K^+K^-)=\frac{1-\dBs}{1-\Ckk\dBs}\left[\Gamma(\Bs(t)\to K^+K^-)\right]_{\dBs=0}\,.
\end{equation}
\item In the untagged rate, besides a global $(1-\Ckk\dBs)^{-1}$ factor, the dominant terms $\SUMCh$ and $\SUMSh$ do not receive significant corrections from $\dBs$ at leading order, and the only potentially sizable corrections appear in $\SUMCc$ and $\SUMSc$.
\item For the CP asymmetry, besides the $(1-\Ckk\dBs)^{-1}$ factor which cancels, the previous comment on the untagged rate does apply to the asymmetry denominator. Similarly, the dominant terms of the numerator, $\DIFCc$ and $\DIFSc$, do not receive significant corrections from $\dBs$ at leading order. The only potentially sizable corrections appear in $\DIFCh$ and $\DIFSh$.
\end{itemize}
Figure \ref{FIG:res:d:I} shows the allowed regions for $\Lkk$ and $\thBs$. Allowing for $\dBs\neq 0$, it is to be noticed that the significance of $\abs{\Lkk}\neq 1$ is lowered from the $3$-$4\sigma$ level to the $2$-$3\sigma$ level. Concerning $\thBs$, while the allowed range for $\im{\thBs}$ is unchanged, larger values of $\re{\thBs}$ are now allowed: $\re{\thBs}=(0.0\pm 4.2)\times 10^{-2}$ and $\im{\thBs}=\left(0.0\begin{smallmatrix}+5.0\\-4.0\end{smallmatrix}\right)\times 10^{-2}$.

\begin{figure}[h!tb]
\begin{center}
\subfigure[$\dBs$ vs. $\re{\thBs}$.\label{sFIG:d:delta:reth}]{\includegraphics[height=0.33\textwidth]{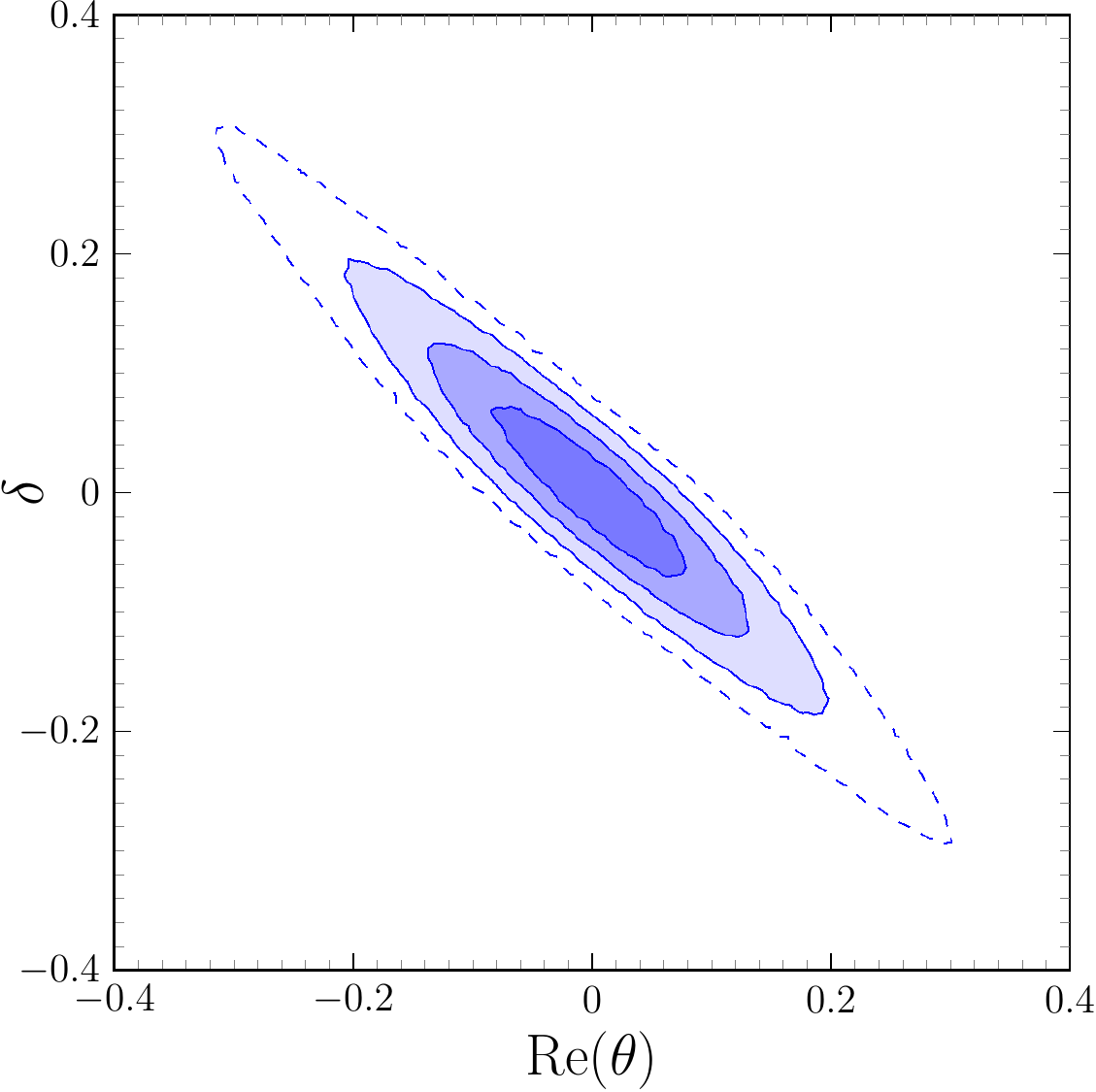}}\qquad
\subfigure[$\dBs$ vs. $\im{\thBs}$.\label{sFIG:d:delta:imth}]{\includegraphics[height=0.33\textwidth]{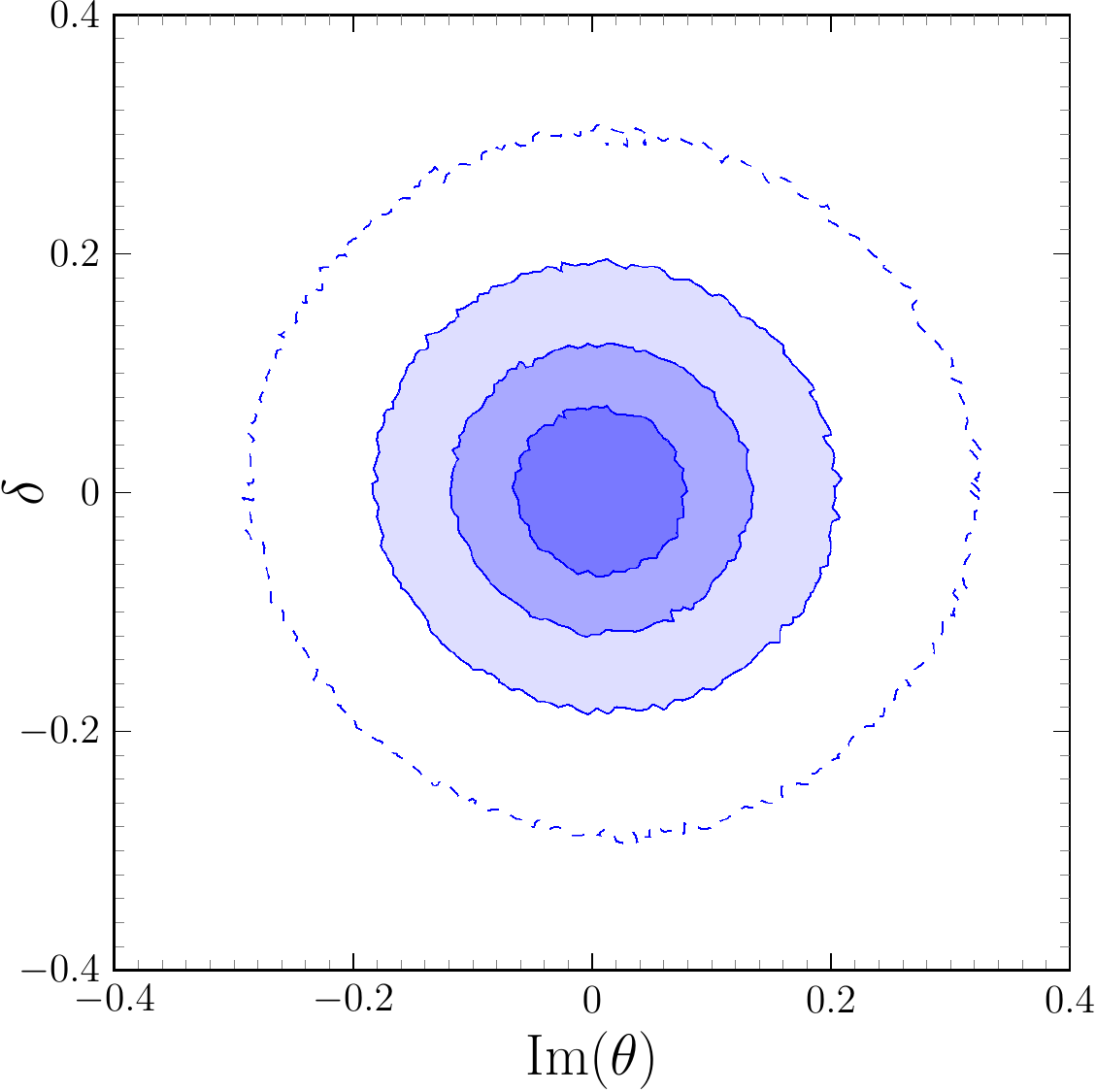}}
\caption{Fit results with $\dBs\neq 0$ (conventions as in Figure \ref{FIG:res:d:I}).\label{FIG:res:d:II}}
\end{center}
\end{figure}

\begin{figure}[h!tb]
\begin{center}
\subfigure[$\Skk$ vs. $\Rkk$.\label{sFIG:SKK:RKK:d}]{\includegraphics[height=0.28\textwidth]{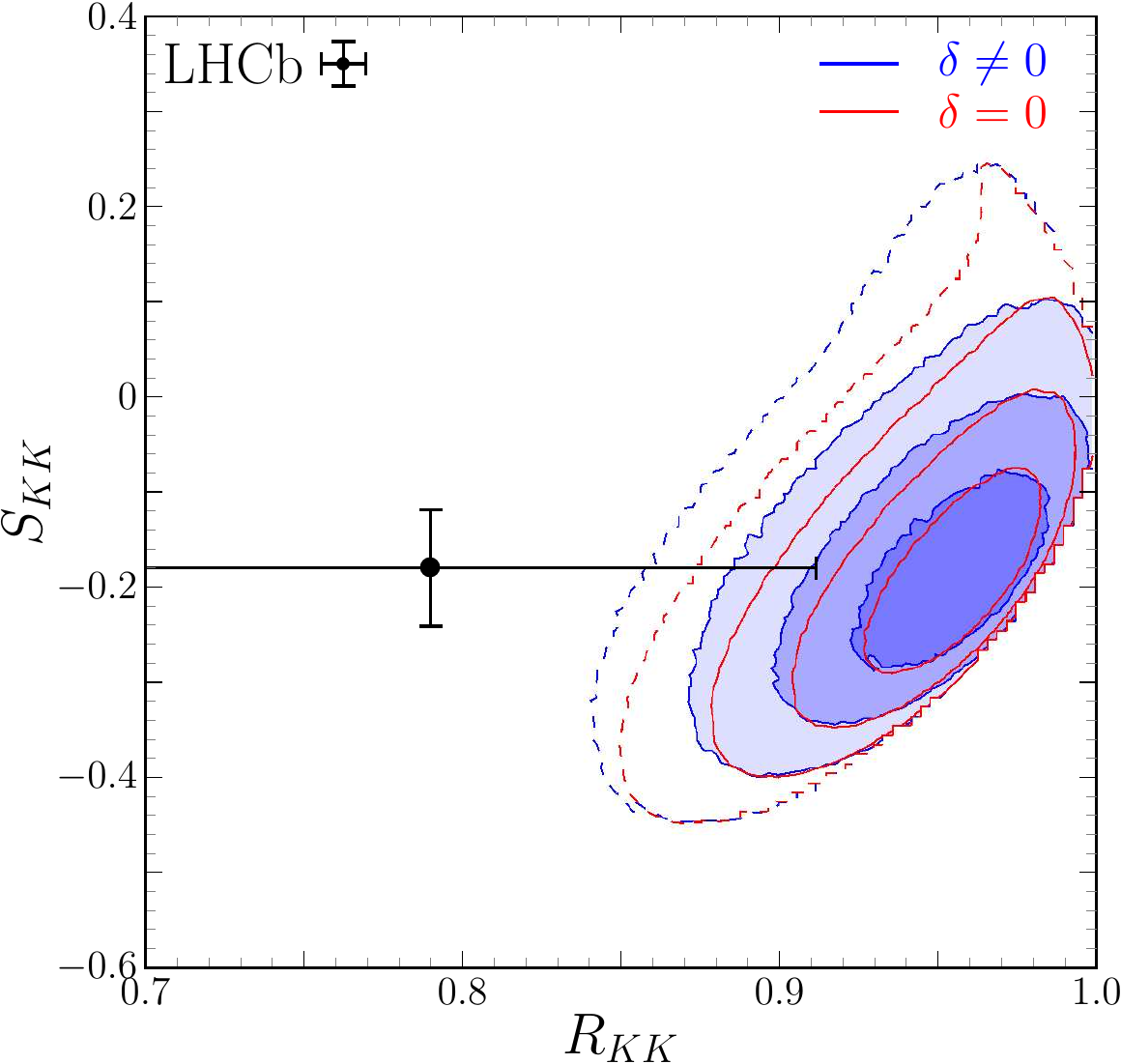}}\quad
\subfigure[$\Ckk$ vs. $\Rkk$.\label{sFIG:CKK:RKK:d}]{\includegraphics[height=0.28\textwidth]{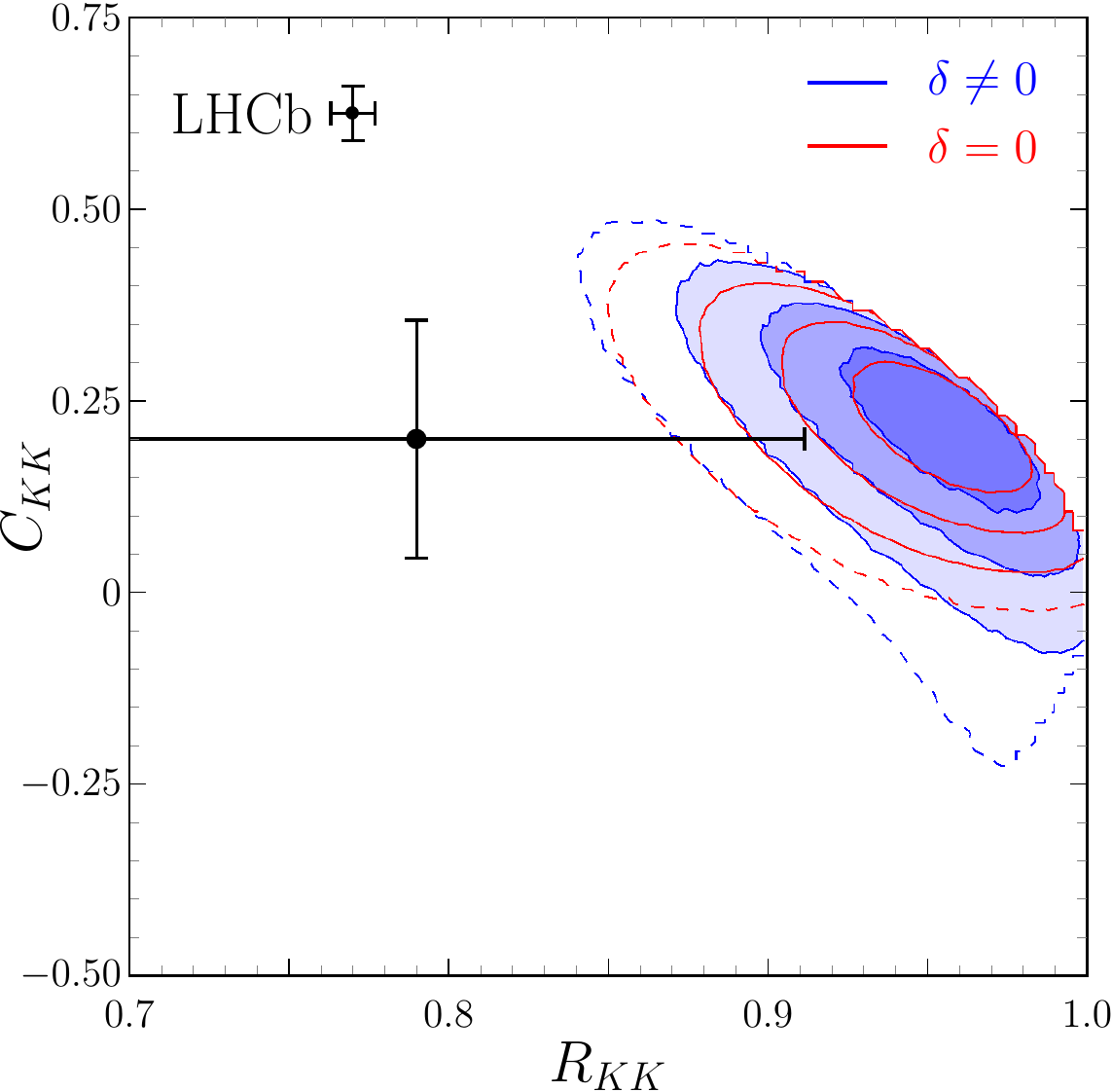}}\quad
\subfigure[$\Skk$ vs. $\Ckk$.\label{sFIG:SKK:CKK:d}]{\includegraphics[height=0.28\textwidth]{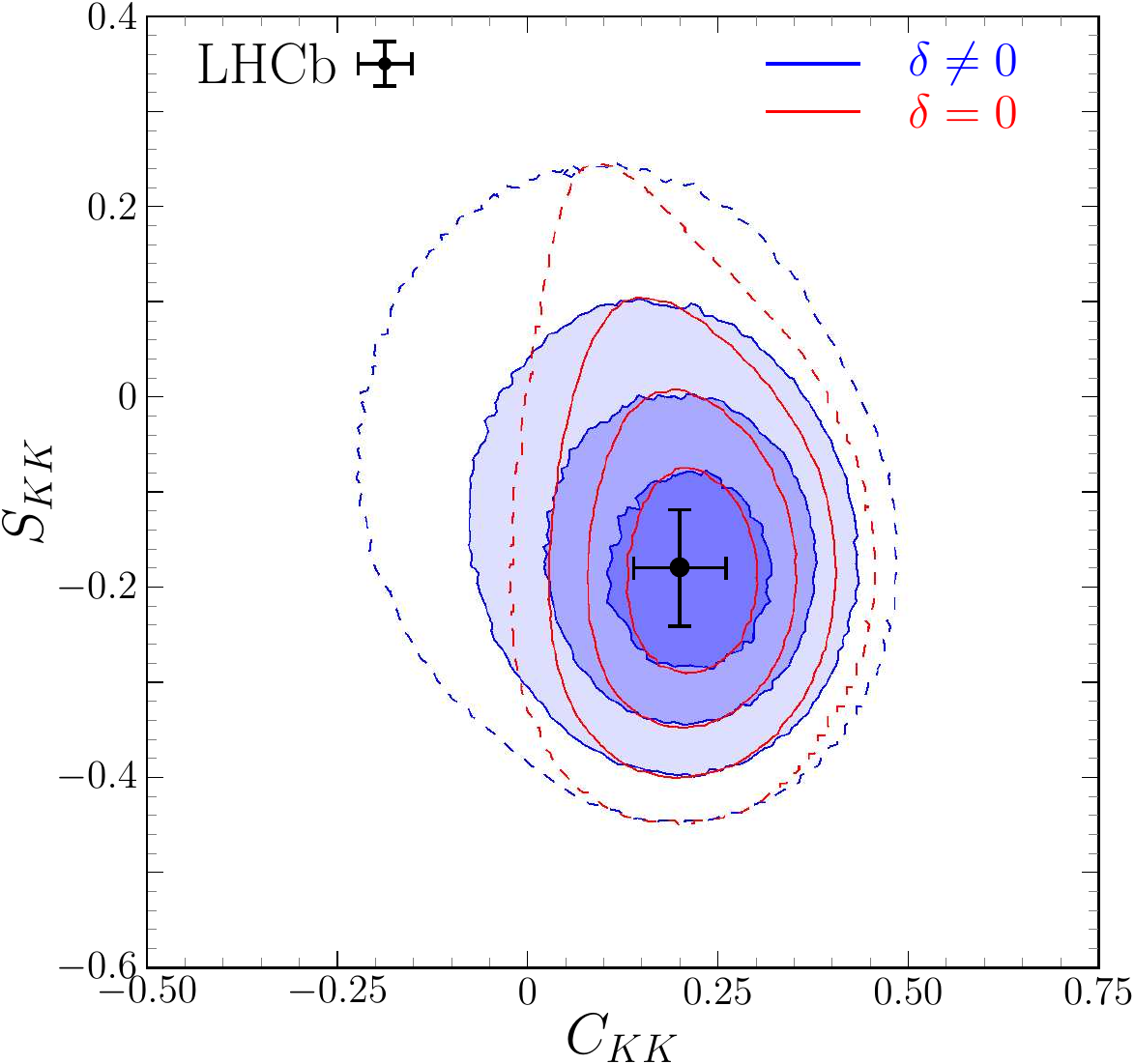}}
\caption{Fit results with $\dBs\neq 0$ (conventions as in Figure \ref{FIG:res:d:I}).\label{FIG:res:d:III}}
\end{center}
\end{figure}

\begin{figure}[h!tb]
\begin{center}
\subfigure[$\dBs$ vs. $\Ckk$.\label{sFIG:d:delta:CKK}]{\includegraphics[height=0.28\textwidth]{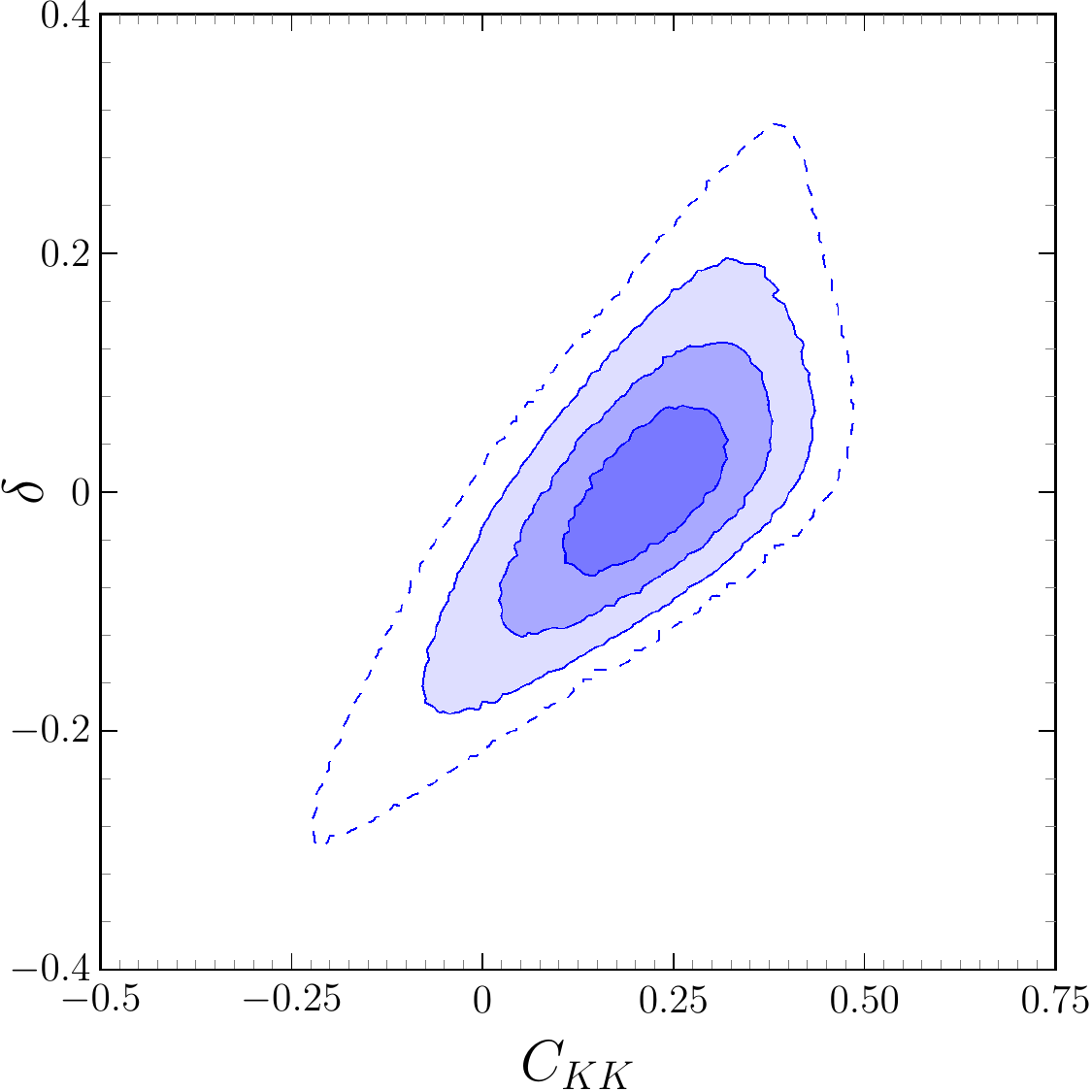}}\quad
\subfigure[$\dBs$ vs. $\Skk$.\label{sFIG:d:delta:SKK}]{\includegraphics[height=0.28\textwidth]{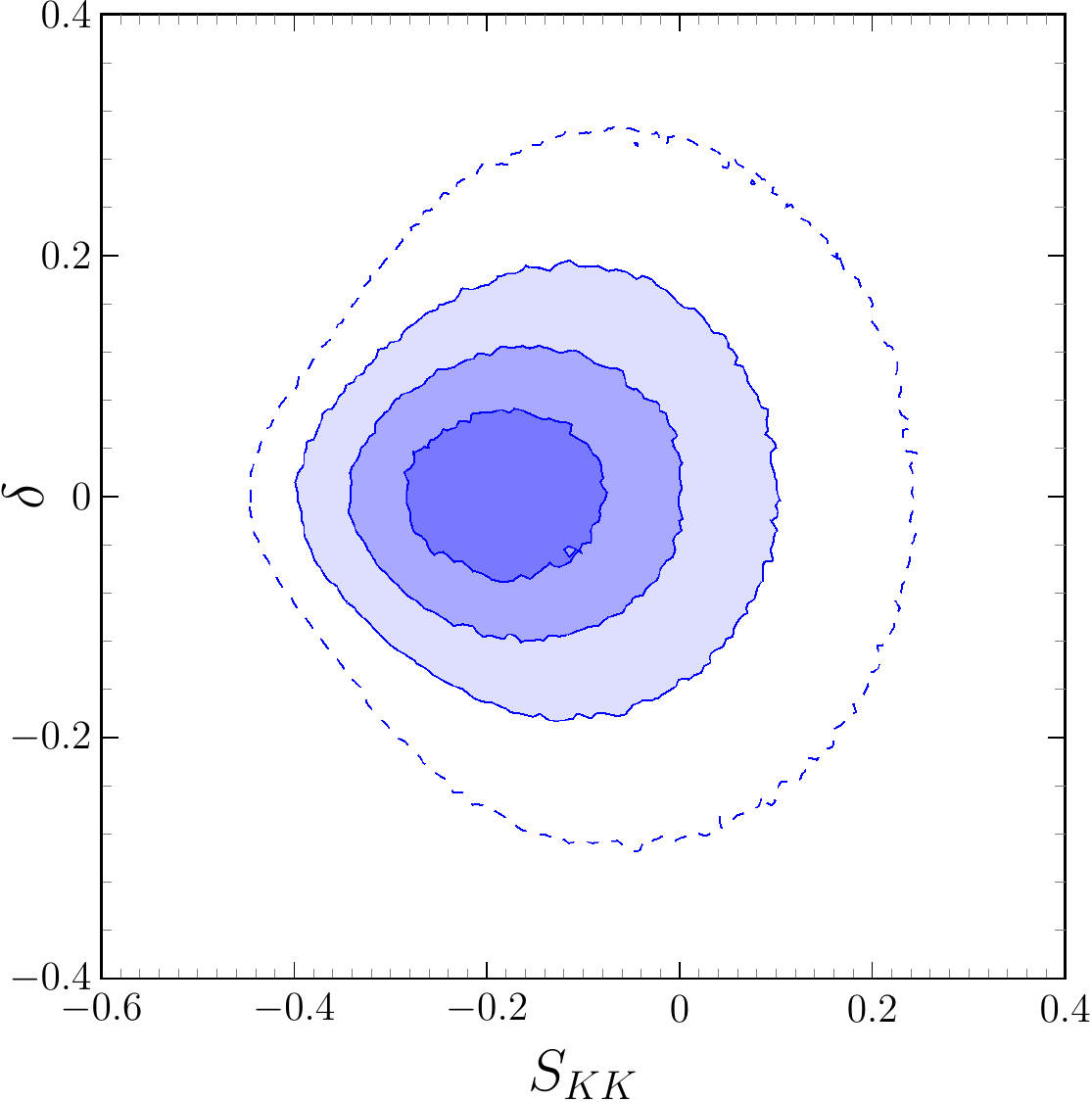}}\quad
\subfigure[$\dBs$ vs. $\Rkk$.\label{sFIG:d:delta:RKK}]{\includegraphics[height=0.28\textwidth]{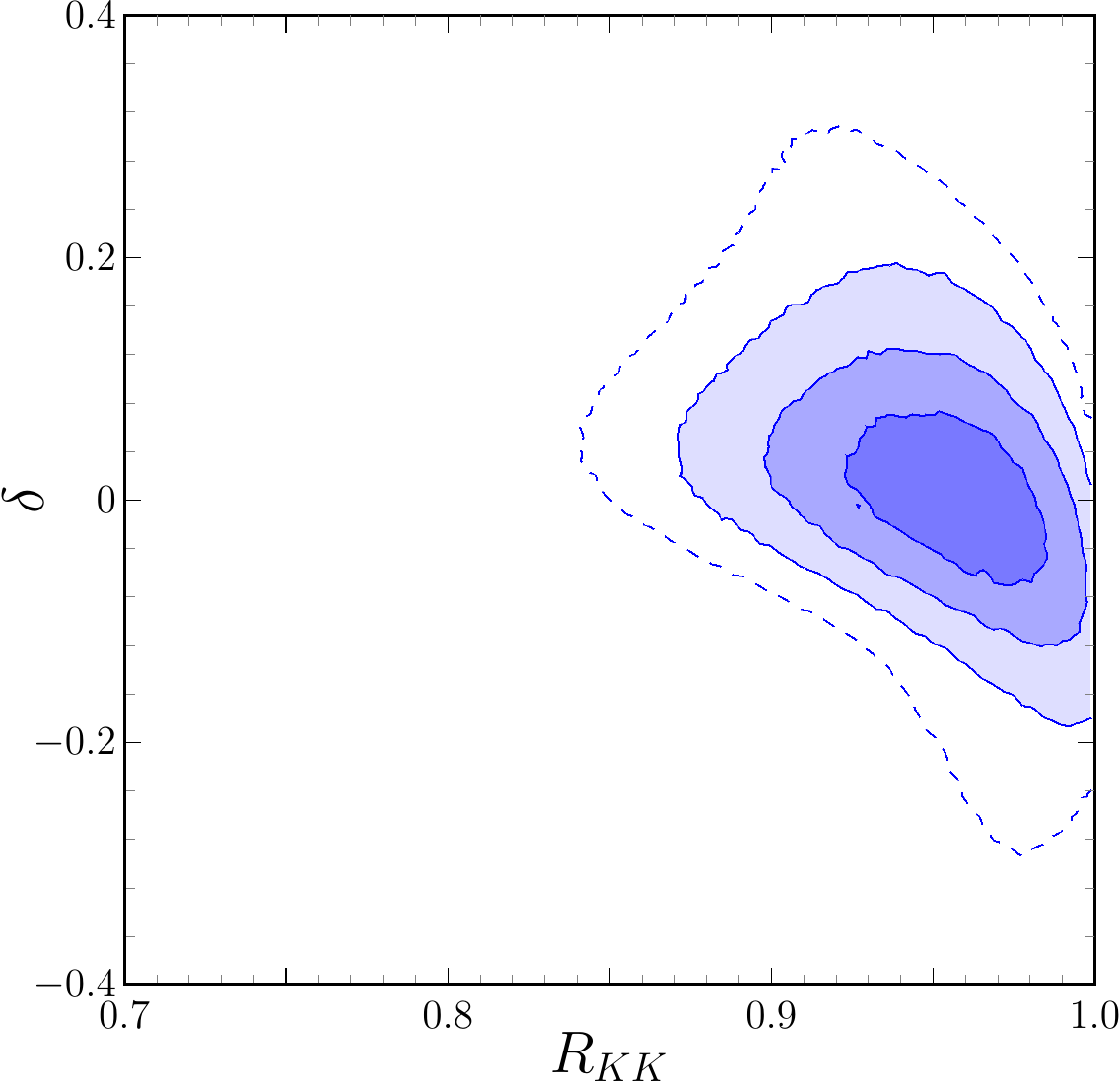}}
\caption{Fit results with $\dBs\neq 0$ (conventions as in Figure \ref{FIG:res:d:I}).\label{FIG:res:d:IIII}}
\end{center}
\end{figure}

\noindent Figure \ref{FIG:res:d:II} shows the allowed regions for $\dBs$ vs. $\re{\thBs}$ and $\dBs$ vs. $\im{\thBs}$. The resulting bounds on $\dBs$, at the $\sim 10^{-1}$ level, are much poorer than existing constraints mentioned above \cite{Aaij:2016yze}. It is also significant, as discussed later, that the allowed ranges for $\dBs$ and $\re{\thBs}$ are strongly correlated.
Figure \ref{FIG:res:d:III} shows the allowed regions for the different pairs in $\{\Ckk,\Skk,\Rkk\}$: while $\dBs\neq 0$ has a minor impact on $\Skk$ and $\Rkk$, it is clear that $\Ckk$ is much more affected, as anticipated by the reduced significance of $\abs{\Lkk}\neq 1$. In terms of $\Ckk$, the significance of $\Ckk\neq 0$ is correspondingly reduced from the $3$-$4\sigma$ level to the $2$-$3\sigma$ level.
Finally, Figure \ref{FIG:res:d:IIII} shows the allowed regions for $\dBs$ vs $\Ckk$, $\Skk$ and $\Rkk$. Following the previous comments, the correlation among sizable values of $\dBs$ and departures of $\Ckk$ from the results with $\dBs=0$ is not surprising.

\noindent As a summary:
\begin{itemize}
\item The analysis with free $\dBs$ does not provide stringent constraints on $\dBs$.
\item While the allowed regions for $\Skk$ and $\Rkk$ do not change significantly (i.e. $\arg(\Lkk)$ is not much affected), there is more allowed space for $\Ckk$ (i.e. $\abs{\Lkk}$ is affected).
\item The enlarged room for $\Ckk$ is correlated with larger allowed values of $\re{\thBs}$. This can be traced back to the sensitivity of the CP asymmetry to the $\DIFCh$ and $\DIFSh$ coefficients in \refEQS{eq:DIFdelta:lin:02}: with $\Rkk\sim 1$, they are sensitive to the combination $\re{\thBs}+\dBs$ rather than to $\re{\thBs}$ and $\dBs$ separately, in agreement with Figures \ref{sFIG:d:delta:reth} and \ref{sFIG:d:delta:CKK}.
\end{itemize}

\clearpage

\providecommand{\href}[2]{#2}\begingroup\raggedright\endgroup

\end{document}